\documentclass{article}
\usepackage{ijcai19}

\usepackage{times}
\usepackage{soul}
\usepackage{url}
\usepackage{hyperref}
\usepackage[utf8]{inputenc}
\usepackage[small]{caption}
\usepackage{graphicx}
\usepackage{amsmath}
\usepackage{booktabs}
\usepackage{xcolor}
\usepackage{multirow}
\usepackage{adjustbox}
\usepackage{array}
\usepackage{pifont}
\usepackage{dblfloatfix}

\title{Advancing Explainable Autonomous Vehicle Systems: \\A Comprehensive Review and Research Roadmap}

\author{
Sule Tekkesinoglu$^1$\footnote{Contact Author}\and
Azra Habibovic$^2$\And
Lars Kunze$^{1}$\\
\affiliations
$^1$Department of Engineering Science\\
  University of Oxford\\
  Oxford, UK \\
$^2$Scania CV \\
\emails
\ sule@robots.ox.ac.uk,
azra.habibovic@scania.com,
lars@robots.ox.ac.uk
}

\begin{document}
\maketitle
\begin{abstract}
Given the uncertainty surrounding how existing explainability methods for autonomous vehicles (AVs) meet the diverse needs of stakeholders, a thorough investigation is imperative to determine the contexts requiring explanations and suitable interaction strategies. A comprehensive review becomes crucial to assess the alignment of current approaches with the varied interests and expectations within the AV ecosystem. This study presents a review to discuss the complexities associated with explanation generation and presentation to facilitate the development of more effective and inclusive explainable AV systems. Our investigation led to categorising existing literature into three primary topics: explanatory tasks, explanatory information, and explanatory information communication. Drawing upon our insights, we have proposed a comprehensive roadmap for future research centred on (i) knowing the interlocutor, (ii) generating timely explanations, (ii) communicating human-friendly explanations, and (iv) continuous learning. Our roadmap is underpinned by principles of responsible research and innovation, emphasising the significance of diverse explanation requirements. To effectively tackle the challenges associated with implementing explainable AV systems, we have delineated various research directions, including the development of privacy-preserving data integration, ethical frameworks, real-time analytics, human-centric interaction design, and enhanced cross-disciplinary collaborations. By exploring these research directions, the study aims to guide the development and deployment of explainable AVs, informed by a holistic understanding of user needs, technological advancements, regulatory compliance, and ethical considerations, thereby ensuring safer and more trustworthy autonomous driving experiences.
\end{abstract}

\section{Introduction}

Autonomous vehicles (AVs) have the potential to enhance road safety, improve traffic flow, reduce carbon footprint, and provide mobility to individuals with limited access. They can reduce road accidents caused by human errors and allow for new forms of time usage and productivity while on the move~\cite{panagiotopoulos2018empirical}. However, widespread acceptance and usage are essential to derive the benefits of AVs. One of the expectations from both consumers and regulators is that AI-driven operations of AVs should be explainable to ensure trust, transparency, and accountability in their decision-making processes~\cite{europaEthicsGuidelines}. To meet this demand, research in the field strives to make AV actions and decisions easily understood and interpreted by humans. AI experts have endeavoured to design models with explainability features and incorporate them into user-centric interfaces, e.g., within the infotainment systems. Numerous literature surveys have scrutinized current approaches from various angles, encompassing cooperative driving, driving scenarios, trustworthiness, and computational infrastructure~\cite{liu2020computing,ren2022survey,jain2022autonomous,kettle2022augmented,kuznietsov2024explainable}. These studies have put forth recommendations and frameworks to pave the way for future investigations that aim to overcome the hurdles associated with developing explainable AVs.

Atakishiyev et al. offered a thorough analysis of eXplainable Artificial Intelligence (XAI) in autonomous driving~\cite{atakishiyev2021explainable}. Their work described concepts and processes involved in AI explainability and proposed a framework that considers societal and legal requirements for ensuring the explainability of AVs. A survey study by Omeiza et al. reviewed research on explanations for various AV operations~\cite{omeiza2021explanations}. The study identified stakeholders involved in developing, using, and regulating AVs. Based on their findings, the authors provided recommendations, including a conceptual framework for AV explainability.

\begin{figure*}[htb!]
\centering
  \includegraphics[width=0.82\textwidth]{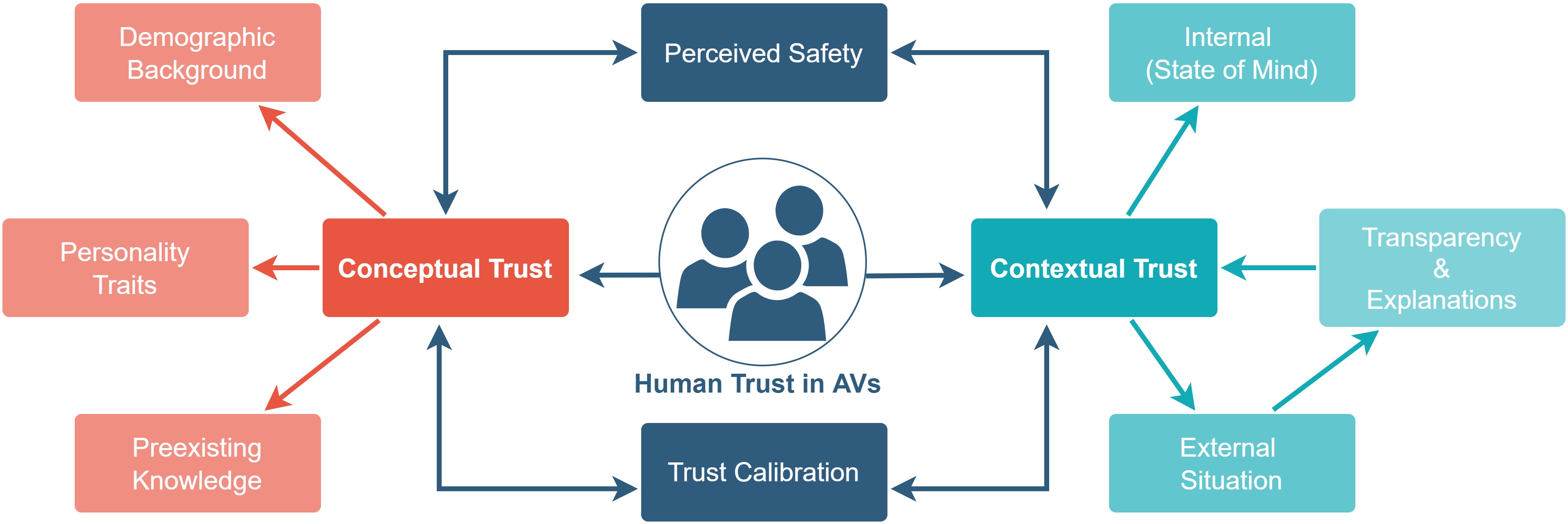}
 \caption{Our perspective on the factors affecting human trust in AVs and its relation to transparency and explanations}
 \label{fig:trust}
\end{figure*}

In their review, Zablocki et al. discuss the challenges of interpretability and explainability in vision-based autonomous driving~\cite{zablocki2022explainability}. Their study provides a detailed organization of the post-hoc explainability methods available for black-box autonomous driving systems. Furthermore, the authors explore several strategies for developing interpretable autonomous driving systems by design. The review by Xing et al. explored human behaviours and cognition in AVs for effective human-autonomy collaboration~\cite{xing2021toward}. The review also delved into the critical factors influencing trust and situational awareness in AVs. Furthermore, the authors provided an analysis of human behaviours and cognition in shared control and takeover situations. In 2022, Kettle and Lee undertook a comprehensive literature review centred on the augmented reality (AR) interfaces implemented in AVs ~\cite{kettle2022augmented}. The review examines the impact of AR interfaces on drivers’ situational awareness, performance, and trust. While their analysis indicates that integrating AR interfaces in AVs can enhance trust, acceptance, and driving safety, the findings underscore the necessity for research on the long-term effects of AR interfaces, impaired visibility, contextual user needs, system reliability, and inclusive design.

Despite the growing body of research on explainability for AVs, it remains unclear how these methods will meet the explanation needs of different stakeholders, such as drivers, passengers, pedestrians, and cyclists. Further analysis is required to determine which situations necessitate explanations and what type of communication is relevant to them. 
The objective of this study is to assess how driving tasks and context influence the explanation needs and identify situations requiring explanations in various levels of automation~\cite{sae2018taxonomy}. Moreover, we evaluate diverse requirements for communicating explanatory information. After critically evaluating the existing literature and identifying research gaps, we propose responsible research and innovation-grounded roadmap for future research\footnote{Responsible Research and Innovation (RRI) is a term used by the European Union's Framework Programmes to describe scientific research and technological development processes that take into account effects and potential impacts on society and the environment.}. With these focus points in mind, our paper contributes in the following ways:

\begin{itemize}
\item	Conducting an extensive survey of recent advancements in explainable autonomous driving.
\item	Presenting factors that affect the explanatory task, information, and information communication.
\item	Offering a roadmap for explainable autonomous driving, grounded in responsible research and innovation, to guide future research in this field.

\end{itemize}

The rest of the paper is organized as follows: Section \ref{sec:back} gives a background on trust factors, transparency, and explanations. Section \ref{sec:meth} presents the methodology describing the design of the review. Section \ref{sec:task} provides a literature review on the explanatory task, followed by explanatory information (Section \ref{sec:info}), and explanatory information communication (Section \ref{sec:com}). Section \ref{sec:road} presents the proposed roadmap and discusses a number of future research directions, while Section \ref{sec:conc} concludes the paper.


\section{Background}
\label{sec:back}
This section lays the groundwork for the review by discussing factors affecting human trust in AV technology and contextualizing its relation to transparency and explanations (See Figure~\ref{fig:trust}). Acceptance and trust depend on an appropriate understanding of the technology. One way to attain this is by transparently elucidating its internal processes. Thus, trust and transparency are intrinsically linked. Transparency fosters confidence in the system by providing insights into its reasoning mechanisms, thereby compelling the necessity for explanations.   

\subsection{Trust in AVs}
\label{sec:trust}

Trust in AVs concerns the reliance, confidence, and belief individuals place in the technology to perform a task accurately, reliably, and safely. \textit{Perceived trust} is crucial in determining how much an individual is willing to accept, intend to use, and depend on the automated system~\cite{panagiotopoulos2018empirical,jardim2013study}. One prominent definition of trust in automation is provided by Lee and See as ``the attitude that an agent will help achieve an individual's goals in a situation characterized by uncertainty and vulnerability''~\cite{lee2004trust}. Several factors influence trust in automation, e.g., perceived safety, design features, brand, road type, and weather. Here, we discuss the factors affecting human trust in AVs under two main concepts: `conceptual' and `contextual' trust (See Figure \ref{fig:trust}).  

Before we delve into these concepts, it is important to point out \textit{perceived safety}'s key role in both conceptual and contextual trust-building processes. Perceived safety could either be objective, based on an objective evaluation of the safety factors, or subjective, that is, safety based on feeling or perception~\cite{li2013study}. Perceived safety is a dynamic process in which levels can shift as individuals encounter new information. For instance, the sudden deviation of AV's driving behaviour from the human driver's norm or even accident reports associated with AVs might hamper the feeling of safety and thereby reduce trust~\cite{yurtsever2020survey}. Therefore, understanding and managing \textit{trust calibration}---the measure of the polarity of trust, that is, misplacement of trust, e.g., overtrust, distrust, and mistrust, is critical~\cite{kraus2020more}. When users perceive the system to perform effectively and reliably, their trust increases---or remains stable. Conversely, users may recalibrate their trust if the system exhibits errors or inconsistencies, potentially becoming more cautious or sceptical. Thus, justified trust calibration is crucial to foster effective collaboration and safe interactions between humans and machines. 


\subsubsection{Conceptual trust}
\label{sec:conceptual}
Conceptual trust refers to a person's mental conception and natural inclination to trust and accept technology over time. This type of trust is influenced by demographic background, pre-existing knowledge, and personality traits. Four main variables that affect trust at the demographic level include culture, age, gender, and education. 
People from different cultures drive by different implicit and explicit rules. For instance,  Edelman et al.'s work reports that participants from different backgrounds respond differently to overtaking and hindering decisions of the AV~\cite{edelmann2021cross}. Though some similarities can be drawn among cultures to form a design framework applicable across the globe, an AV is not expected to have the same behaviour worldwide and be understood and accepted by everyone equally well. Cross-cultural differences require AVs in different places to adapt to the culture and transition from one behaviour to another when it travels through cultural borders~\cite{pillai2017virtual}.

Regarding the age factor, research revealed no significant differences between the average trust ratings of younger and older adults~\cite{rovira2019looking,liu2019effect,hartwich2019first}. Both younger and older participants reported autonomous driving as trustworthy and acceptable. Liu et al. reported that the younger participants had a higher positive attitude and acceptance than the older participants~\cite{liu2019effect}. At the same time, Hartwich's study revealed that older people also had a strong positive attitude toward AVs~\cite{hartwich2019first}. Further research is needed to assess if consistent age differences exist in individual trust in AVs. In terms of gender differences, previous studies suggest that men are more inclined towards accepting AVs~\cite{zoellick2019amused,qu2019development}. Regarding the education level, Liu et al. showed that education was positively correlated with respondents' willingness to pay for AV technology~\cite{liu2019effect}.
                           
On another note, research has explored how personality traits, particularly the big five, influence the acceptance of AVs. People with high openness, conscientiousness, extraversion, and agreeableness had a more positive perspective, while high neuroticism had a negative effect~\cite{qu2021effects,irfan2021relating}. Personal innovativeness, considered a relevant personality trait influencing one's willingness to accept and implement new ideas, products, and systems~\cite{ali2019personality}, is also perceived to positively influence the adoption of AVs~\cite{hegner2019automatic}.

Pre-existing knowledge is another factor influencing trust formation even before an actual interaction with AVs occurs. Prior experience with similar technology---or other transport modes and information conveyed by external sources (e.g., media) could lead to overtrust or distrust toward AVs~\cite{papadimitriou2020transport,velasco2019studying}. Previous studies reported that participants with pre-existing knowledge of AVs held more optimistic views towards using AVs in the future than those without such knowledge~\cite{kaye2020priori}. Research by Shammut et al. also highlighted that respondents who tended to trust AVs commented on how safe and reliable AVs are based on their pre-existing knowledge~\cite{shammut1aautonomous}. Another work revealed that responders who reported greater knowledge and experience of new technologies were more accepting of AVs~\cite{lee2017age}. On the other hand, the study by Zhou et al. revealed that misinformation could be the most significant factor causing distrust in AVs due to risks associated with software malfunction or cyberattacks~\cite{zhou2021factors}. According to Zhou et al., negative news coverage about accidents involving AVs could cause people to distrust them~\cite{zhou2023examining}. While the level of trust in AVs can be related to pre-existing knowledge and previous experiences with technology, trust can be gradually developed after interacting with AVs.


\subsubsection{Contextual trust}

Contextual trust is determined by the particular situation or location in which an interaction occurs. Contextual trust is consequential and temporal, influenced by both the \textit{external} (e.g., driving behaviour) and \textit{internal} environment (e.g., state of mind)~\cite{kearns2023contextual}. People might feel positive trust toward a system concerning certain tasks and goals yet mistrust or distrust when other tasks, goals, or situations are involved. 
Existing literature shows contextual elements, such as traffic signals and driving behaviour (i.e., defensive, normal, and aggressive), are important external factors related to contextual trust in pedestrian-AV interaction~\cite{jayaraman2019pedestrian}. Research showed that pedestrians generally expressed more trust in AVs at signalized crosswalks. This finding is based on the assumption that AVs are expected to be law-abiding (e.g., stopping at red lights) under all circumstances~\cite{millard2018pedestrians}.

In future, human-driven vehicles and AVs will inevitably share the road. A recent work studied how human drivers' trust toward AVs varies with contextual attributes such as traffic density and road congestion. Sun et al. discovered a positive correlation between the vehicle gap and trust level, while factors such as relative speed and traffic density exhibited a negative correlation~\cite{sun2023modeling}. Osikana et al. found that the vehicle's speed and distance have the most influence on the cyclists' trust levels~\cite{oskina2023safety}. Another study found that motorcyclists trust AVs more than human drivers due to unpredictable driving behaviours, not checking their blind spots, and speeding~\cite{pammer2023humans}. 

Researchers have investigated the influence of perceived risk on drivers' trust in automated driving. Ayoub et al. examined drivers' trust in AVs' takeover scenarios with different system performances~\cite{ayoub2021investigation}. The results show that drivers quickly calibrate their trust level when AV clearly demonstrates its limitations. Azevedo-Sa et al. considered two risk types (i.e., low system reliability and low visibility from foggy weather) as contextual factors where low visibility did not significantly impact drivers' trust in automated driving~\cite{azevedo2021internal}.

Trust develops inconsistently in different contexts due to \textit{internal factors} such as emotional state, attentional capacity, and self-confidence. The attentional capacity of a driver often depends on the task---monotonous automated driving promotes passive task-related fatigue~\cite{korber2014potential}. A high level of positive trust in automation might reduce attention allocation and situational awareness. Inactivity and engagement in non-driving-related tasks have been suggested to impair the driver’s ability to handle the takeover process safely, which might cause accidents resulting from over-trusting automation~\cite{naujoks2018partial}. In another work, contextual trust was associated with emotions, where a high level of trust in AVs significantly improved participants’ positive emotions~\cite{avetisian2022anticipated}. Furthermore, factors such as motivation, stress, sleep deprivation, and boredom influence trust dynamics in driving scenarios~\cite{hoff2015trust}.

A non-task-related condition, attention deficit hyperactivity disorder (ADHD), could influence trust in AVs due to the cognitive challenges it poses in monitoring and maintaining awareness~\cite{korber2014potential}. Since people with ADHD are more prone to distraction, their ability to monitor automation and respond appropriately might be compromised, further impacting their willingness to use automation. On a different note, self-confidence is another context-dependent variable that can alter trust in automation and control allocation~\cite{hoff2015trust}. A recent study compared participants who underwent training for engaging automated driving systems with those who did not receive such training. The findings unveiled notable disparities in response to emergency events requiring drivers to assume control. Trained drivers demonstrated reduced reaction times and exhibited more calibrated trust levels in automation compared to their non-trained counterparts~\cite{payre2017impact}.

As discussed above, some factors influencing trust in AVs may not be addressed by training or technological advancements. Regardless, certain elements are central to establishing trustworthiness objectively, including technical competence, adept situation management, and transparency~\cite{choi2015investigating,nastjuk2020drives}. Technical competence refers to the user's belief that the AV meets performance and reliability expectations (e.g., GPS connection, security, and ability to choose the best course of action). Situation management relates to the user's belief that they can gain control over the vehicle or contact a human operator whenever required. Transparency provides a clear view of AVs abilities and operations to promote appropriate trust and discourage automation misuse and disuse. In the following, we delve into the transparency requirement further.

\begin{table*}[!b]
\centering
\caption{Research questions under the topics of explanatory tasks, information, and information communication}\label{tab:Rqs}

\begin{tabular}{lll}
\hline
Research Topics                                                                         &  & Research Questions                                                         \\ \cline{1-1} \cline{3-3} 
\multirow{2}{*}{\begin{tabular}[c]{@{}l@{}}Explanatory \\  Task\end{tabular}}              &  & \hyperref[sec:RQ1]{RQ1.} How do task and context influence the need for an explanation?              \\
                                                                                           &  & \hyperref[sec:situations]{RQ2.} What are the timings and situations requiring an explanation?                     \\ \hline
\multirow{2}{*}{\begin{tabular}[c]{@{}l@{}}Explanatory \\ Information\end{tabular}}        &  & \hyperref[sec:RQ3]{RQ3.} What are the layers of transparency for the explanatory information?  \\
                                                                                           &  & \hyperref[sec:RQ4]{RQ4.} How do task and context influence the explanatory information?        \\ \hline
\multirow{2}{*}{\begin{tabular}[c]{@{}l@{}}Exp. Information \\ Communication\end{tabular}} &  & \hyperref[sec:RQ5]{RQ5.} How is the explanatory information communicated to internal stakeholders?\\
                                                                                           &  & \hyperref[sec:RQ6]{RQ6.} How is the explanatory information communicated to external stakeholders?\\ \hline
\end{tabular}
\end{table*}

\subsection{Transparency}

System transparency refers to the degree of observability and predictability of the operations to give a sense of control and acceptance in autonomous driving~\cite{nastjuk2020drives}. Transparency is usually a precondition for \textit{accountability} in which AV should be able to explain its plans and actions, especially in unexpected events. It concerns the extent to which the responsibility for the outcome can be ascribed to an agent (e.g., governments, companies, system developers, drivers) legally or ethically~\cite{vilone2020explainable}. This is because, for an action to be evaluated properly, relevant stakeholders should have access to all necessary information.  

Transparency requirements vary depending on the levels of autonomy (e.g., SAE level 3 vs level 5)~\cite{sae2018taxonomy}. In semi-autonomous systems, transparency is required when the vehicle needs to hand the control to the driver at certain stages of the journey. In the highest level of vehicular automation (level 5), where vehicles can handle all traffic situations, transparency requirements may not necessarily be safety-critical, but in the attempt to increase trust in the automation, thereby improving user experience and comfort~\cite{diels2018information,elbanhawi2015passenger}. Regardless of the level of autonomy, AVs in action are expected to communicate information about the driving decisions, reasons, and plans transparently in a timely manner to calibrate an appropriate level of trust~\cite{xing2021toward}. 

\begin{figure*}[htb!]
\centering
  \includegraphics[width=0.8\textwidth]{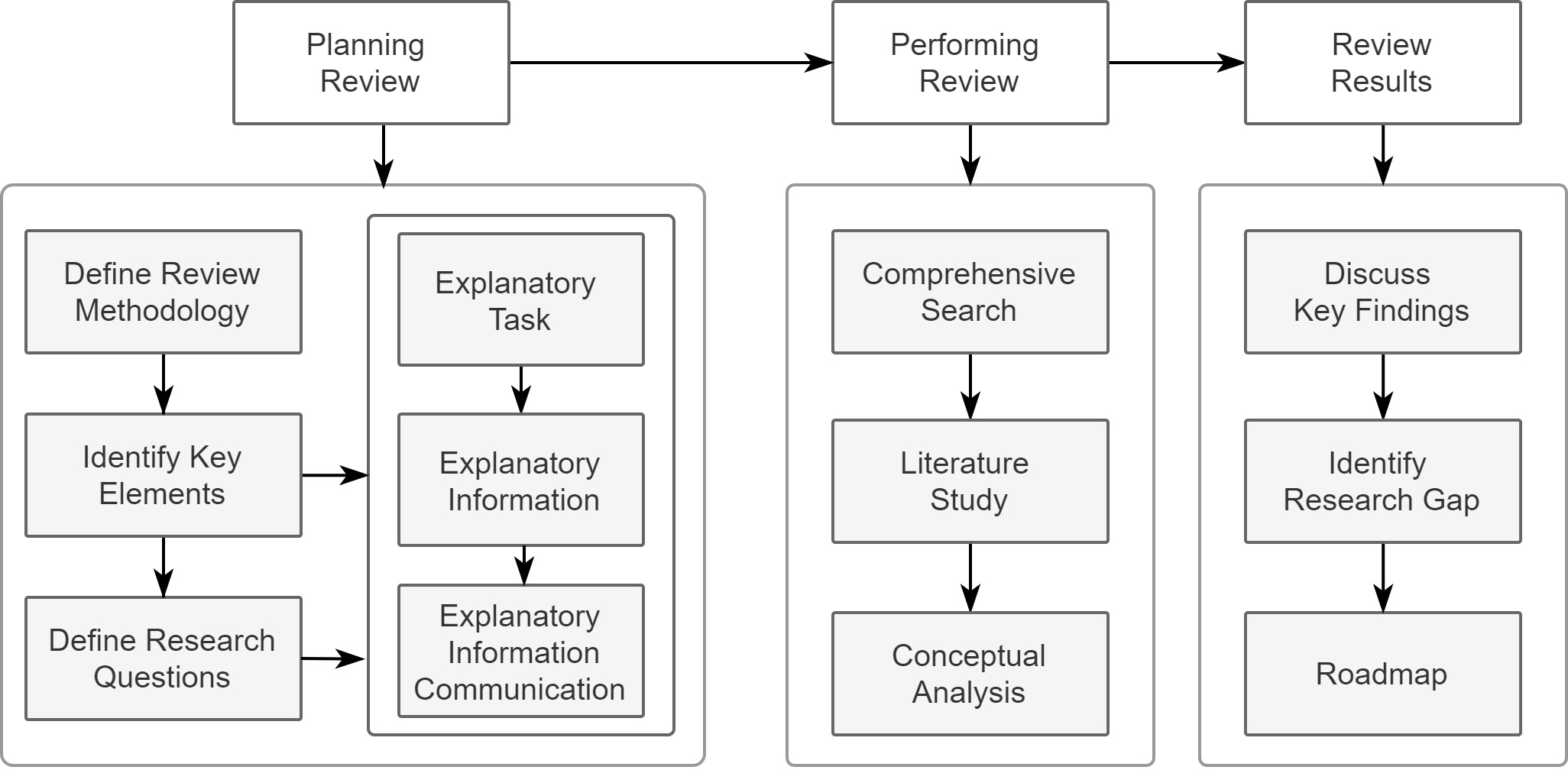}
 \caption{The review methodology}
 \label{fig:review}
\end{figure*}

\subsection{Explanations}

Explanations are a mechanism to promote greater transparency and help address problems caused by the black-box nature of automated decision-making. The literature has extensively studied explanations for AV’s behaviour to improve transparency of system decisions and trust. The research explored transparency and explanations for situations including system uncertainty~\cite{helldin2013presenting}, perceived risk (e.g., weather and driving speed)~\cite{ha2020effects}, intent communication for vulnerable road users~\cite{dey2018interface}, motion planning~\cite{gyevnar2022human}, driver takeover request~\cite{korber2018introduction}, and lane positioning~\cite{kim2018textual}. 

Our review found two main perspectives in the context of explanations for AVs: `user-centric' and `technical-centric' approaches.  User-centric explanations prioritize conveying information to users (i.e., the general public or non-experts) in a comprehensible and meaningful manner. These explanations involve intuitive visualizations or simple language to communicate complex technical decisions. User-centric explanations centre around the driving context and the kind of information or interaction that is relevant. Contexts (e.g., near-crash) and individual attributes (e.g., aggressive driving style) significantly influence the desire for explanation. A study showed that people tend to agree on the need for an explanation for near-crash or emergency driving scenarios and less for the ordinary driving situation~\cite{shen2020explain}. 
The context also has a strong influence on the types of questions the AV may receive. Omeiza et al. explored explanation types (i.e., causal and non-causal) and their investigatory queries (e.g., why, why not, what if, how), demonstrating `why-not' (i.e., contrastive) explanations have the most positive impact on participants' understandability~\cite{omeiza2021towards}. 

On the other hand, technical-centric explanations delve into the intricate details of how the AV operates, investigating technical aspects, algorithms, or sensor data used in decision-making. These explanations are more geared towards experts or individuals with technical backgrounds who seek in-depth insights into the underlying mechanisms of the AV's actions. Technical-centric explanations enable experts to adjust an underlying model to improve accuracy and efficiency. 
Often, a combination of both user and technical-centric approaches might be necessary to address various user needs and levels of expertise.

\section{Review Methodology}
\label{sec:meth}

This study demonstrates a review of the design considerations for explainable AV. We adopt the `state-of-the-art review' methodology, in which we analyze the subject matter in contrast to current and other combined retrospective approaches, offering a new perspective (See Figure \ref{fig:review}). In order to map out the existing literature, we identified three research topics: explanatory tasks, explanatory information, and explanatory information communication. Within each topic, we have delineated multiple research questions (See Table \ref{tab:Rqs}) that are further explored in Sections \ref{sec:task}, \ref{sec:info}, and \ref{sec:com}.





Given that the scope of this study is limited to the explainability of AVs, it does not encompass the broader field of explainable AI research. To obtain relevant articles, we applied the keywords `explainable + autonomous + vehicles + driving' and `transparency + feedback + communication + autonomous + vehicle + car + driving' on Google Scholar for publications between 2018 and 2023. Our initial search yielded 169 papers; through the review process, we eliminated papers that lacked a component of explainability/transparency, only presented a conceptual framework, or were duplicates. This left us with 135 relevant papers. After critically evaluating these papers and identifying research gaps, we formulated a conceptual roadmap for future research tackling advancing explainable AV systems. 

The roadmap is grounded in a responsible research and innovation (RRI) framework. RRI frameworks provide guidelines and good practices to promote inclusive and sustainable research and innovation design, proactively addressing challenges before they arise when deploying AVs in the broader society~\cite{burget2017definitions}. The roadmap presented here follows the AREA framework, which stands for \textbf{A}nticipate, \textbf{R}eflect on, \textbf{E}ngage with, and \textbf{A}ct upon~\cite{ukriFrameworkResponsible}. Each holds a specific role in the research and innovation process, as follows:

\textbf{Anticipate:} Conducting an in-depth analysis of the intended or unintended consequences that may arise. This aims to facilitate the exploration of potential implications that may otherwise remain overlooked and under-discussed.

\textbf{Reflect:} Reflecting on the objectives, motivations, and potential implications of the research, along with any unknowns, gaps in knowledge, presuppositions, dilemmas, and societal shifts that may arise.

\textbf{Engage:} Encouraging diverse perspectives and inquiries to be shared and discussed in a collaborative and inclusive manner.

\textbf{Act:} Utilizing these conversations to shape the course of the research and innovation process itself.

Taken together, we propose potential areas of improvement related to each research topic and then discuss these further within the AREA framework in Section \ref{sec:road}.

 \section{Explanatory Task}
 \label{sec:task}
Explanations serve different purposes in different settings. The task and context at hand shape the necessity for an explanation. We reviewed pertinent literature to identify circumstances necessitating explanation---addressing research questions RQ1-RQ2. 
Table \ref{tab:compare} offers a comprehensive overview of the literature with regard to factors influencing the explanatory task. These have been qualitatively evaluated using four levels of assessment: addressed, partially addressed, barely addressed, and not addressed. The following section provides an in-depth analysis of the literature through the lenses of these factors.

\begin{table*}[htbp]
\centering
\caption{Qualitative assessment of the literature in terms of factors influencing explanatory task}\label{tab:compare}
 \begin{adjustbox}{width=0.9\textwidth}

\begin{tabular}{lllllllllllllll}
\hline
Autonomy &  & \multicolumn{3}{l}{\hspace{0.7cm}Stakeholder}               &  & \multicolumn{4}{l}{\hspace{1.2cm}Driving Operations}          &  & \multicolumn{4}{l}{ \hspace{1.2cm}Timing and Situation Criticality}                                                                                                                                    \\ \cline{1-1} \cline{3-5} \cline{7-10} \cline{12-15} 
           &  & {\rotatebox[origin=c]{60}{Internal}} & {\rotatebox[origin=c]{60}{External}} & {\rotatebox[origin=c]{60}{Observations}} &  & {\rotatebox[origin=c]{60}{Perception}} & {\rotatebox[origin=c]{60}{Planning}} & {\rotatebox[origin=c]{60}{Localization}} & {\rotatebox[origin=c]{60}{Control}} &  & {\rotatebox[origin=c]{60}{Proactive critical}} & \begin{tabular}[c]{@{}l@{}}{\rotatebox[origin=c]{60}{Proactive non-critical}}\end{tabular} & {\rotatebox[origin=c]{60}{Reactive critical}} & \begin{tabular}[c]{@{}l@{}}{\rotatebox[origin=c]{60}{Reactive non-critical}}\end{tabular} \\ \hline
Level 1    &  & \ding{109}        & \o{}        & \o{}                       &  & \ding{109}          & \o{}        & \o{}            & \ding{109}       &  & \o{}                  & \ding{109}                                                                       & \o{}                 & \o{}                                                                      \\ \hline
Level 2    &  & \ding{109}        & \o{}        & \o{}                       &  & \o{}          & \o{}        & \o{}            & \ding{109}       &  & \ding{109}                  & \o{}                                                                       & \o{}                 & \o{}                                                                      \\ \hline
Level 3    &  & \ding{119}      & \o{}        & \o{}                       &  & \o{}          & \ding{119}      & \o{}            & \ding{119}     &  & \ding{119}                & \ding{109}                                                                       & \ding{109}                 & \o{}                                                                      \\ \hline
Level 4    &  & \ding{108}       & \ding{119}      & \o{}                       &  & \ding{119}        & \ding{119}      & \ding{119}          & \ding{108}      &  & \ding{119}                & \ding{109}                                                                       & \ding{109}                 & \ding{109}                                                                      \\ \hline
Level 5    &  & \ding{108}       & \ding{108}       & \o{}                       &  & \ding{119}        & \ding{119}      & \ding{119}          & \ding{108}      &  & \ding{119}                & \ding{119}                                                                     & \ding{109}                 & \ding{109}                                                                      \\ \hline
\end{tabular}
\end{adjustbox}
\\[5pt]
\ding{108} addressed, \ding{119} partially addressed, \ding{109} barely addressed, \o{} not addressed  
\end{table*}

\subsection{RQ1: How do task and context influence the need for an explanation?} 
\label{sec:RQ1}
The need for explanation varies depending on the stakeholders, driving operations, and the level of autonomy. 

\subsubsection{Stakeholders}
\label{sec:user}
Within the autonomous driving domain, explanations are utilised for multiple purposes, resulting in various types of stakeholders. The level of detail, form, and mode of communication necessary for an explanation vary depending on the target audience and the explanation's purpose. As such, it is vital to tailor the explanation to meet the specific needs of the intended stakeholders. Omeiza et al. have identified three primary stakeholder categories: Class A (including all end-users and society), Class B (consisting of technical groups such as system developers), and Class C (encompassing regulatory bodies, such as guarantors)~\cite{omeiza2021explanations}.

We further divided Class A into two main classes of users: vehicle internal and external stakeholders. Internal stakeholders include drivers who participate in the driving operations and passenger(s) who may interact with the AV but are not responsible for any driving operation.  External stakeholders, on the other hand, are remote operators who can assist or drive the AV, vulnerable road users, including pedestrians, cyclists, motorcyclists, and mobility scooter users, as well as protected road users such as drivers of cars, trucks, and buses~\cite{hollander2021taxonomy}. 
The results revealed that only a fraction of the papers---67 out of 135---specified the intended stakeholder for the explanations. While the rest of the papers did not mention the target stakeholder, it was evident that these studies were intended for technical users to explain model predictions. Technical-centric explanations are also relevant to Class A users; however, further research is necessary to study how these approaches can be adapted to meet the end-user requirements and integrated into appropriate human-machine interfaces (HMIs). Moreover, previous studies have not yet touched upon integrating stakeholder observations/feedback into explanations.


\subsubsection{Driving operations}

Another important factor that influences the explanatory task is the driving operation. AVs, equipped with sensory capabilities to perceive their surroundings, rely on a series of interconnected operational stages to make informed driving decisions in real-time. These stages encompass perception, localization, planning, and control. 

Perception is the sensing of an operational environment defined as a combination of two tasks: road surface extraction and on-road object detection through sensors such as vision, LiDAR, RADAR, and ultrasonic technology. Gradient-based methods are the most commonly used approaches for scene understanding tasks~\cite{mankodiya2022od,abukmeil2021towards,kolekar2022explainable}. Various explainability approaches have been developed for early anticipation of traffic accidents~\cite{monjurul2021towards}, risk assessment in complex road conditions~\cite{yu2021scene}, and segmentation under hazy weather~\cite{saravanarajan2023improving} based on the vision data. Explanations are explored for LiDAR and RADAR data to detect 3D objects~\cite{pan2020towards} and sensor input uncertainty for possible deceptive attacks~\cite{rastogi2022explaining}. Other works have proposed explainable localization and mapping by extracting non-semantic features from LiDAR scans to accurately determine AV's position in the environment~\cite{wober2020autonomous,charroud2023xdll}.    

As the AV perceives its surroundings and gets its precise localization, it plans the trajectory from the initial point to the final destination. Planning is a complex operation in that the amount of data the AV processes per time is hard to keep track of continuously and accurately. Research in this field has focused on either route planning (i.e., selection of a route)~\cite{limeros2022towards,renz2022plant,albrecht2021interpretable} or behaviour planning, which involves anticipating and interacting with other road users who share the same trajectory~\cite{zhang2023keeping,zhang2022trajectory,lee2019joint,bouchard2022rule}. Much of this research has focused on~\textit{explicable planning} models in which the outcomes can be explained but not easily understandable by general users. The process requires further translating the AV’s updated plans into comprehensible and user-friendly formats.

Finally, control of an AV involves the execution of planned motions. Feedback controllers mainly manage this function by interacting with the sensors and assisting the car in controlling its trajectory along the journey. The review results show that explanations are generated and explored mainly for the control and navigation-related tasks, i.e., longitudinal control (speed regulation) and lateral control (steering operation). Having that the driving operations are interconnected, the explanation generation process for control and navigation operation is coupled with perception and planning tasks. Explanations are explored for each and every driving action, such as move, stop, and lane change in urban areas~\cite{omeiza2023effects,jing2022inaction}. From the technical perspective, the aim of generating explanations for each action taken by AV is to understand how the model would perform under unseen environments and unexpected situations~\cite{feng2023nle,kochakarn2023explainable,ben2022driving}. From the user-centric perspective, explaining every action helps build the user's initial trust and acceptance that the AV is making reasonable decisions aligned with their expectations. In summary, our review reveals that research has primarily focused on control and perception tasks, particularly for higher-level automation while lacking in-depth studies for explanations for planning and localisation tasks.

\subsubsection{Function of explanations in different levels of autonomy}

It is important to consider the level of autonomy inherent in the system to understand the function of explanations. Vehicles equipped with higher autonomy may necessitate fewer explanations in certain scenarios, while those with lower autonomy may demand more. This discussion entails delineating the situations requiring explanations based on the level of autonomy, whether it involves collaborative driving or full autonomy. 

In low-level automation (SAE Level 1-3), the driver has the responsibility of keeping an eye on the driving environment, executing either longitudinal (accelerating, braking) or lateral (steering) dynamic driving tasks, and monitoring the performance of the driver assistance system. In this level of autonomy, explanations are required for the driver where they need to interact with the AV within the shared control context. The function of explanations in various scenarios within autonomy levels 1-3 can be outlined as follows:

\begin{description}

\item[] \textbf{Maintaining situation awareness--} Maintaining situation awareness is essential for the driver to take control at any point of the journey. The driver continuously monitors the driving behaviour and makes necessary adaptations if the current driving behaviour differs from their desired actions~\cite{wiegand2020d}. Explanation mechanism keeps the human driver informed in case of system uncertainty~\cite{kunze2019automation,kunze2019function,kruger2020feeling}, perceived risk~\cite{malawade2022roadscene2vec,nahata2021assessing,chaczko2020exploration,candela2023risk,stocco2022thirdeye}, and traffic complexity~\cite{hartwich2021improving,albrecht2021interpretable}. 

\item[] \textbf{Safe transition to manual driving--} In complex traffic scenarios, it is possible for a driver to miss a road hazard. In such cases, a takeover request without context may not effectively convey the gravity of the situation, and the driver may take over the control without fully comprehending the situation. This can lead to a decrease in the quality of takeover control~\cite{zhang2023keeping}. A study by Chen et al. suggests that situation influences participants' decisions to take over control when observable cues are not available in the driving environment~\cite{chen2023adding}. It is necessary to provide clear explanations to persuade the driver to take the necessary action to avoid potential hazards, e.g., the system sends an alerted message: ``Automated driving is about to be disabled due to the busy intersections, switch to manual driving mode and proceed with caution''~\cite{naujoks2019towards,du2021designing}.   

\item[] \textbf{Supporting trust calibration--} Another function of explanations is to convey the AV's performance limits to support trust calibration~\cite{kunze2018augmented,kunze2019conveying}. Explanations provide a clear view of AV's abilities and operations to promote appropriate trust and discourage misuse of automation by over-trusting it or disusing it due to under-trusting it. Goldman and Bustin show that the explanation presenting risk and the next plan reduces participants' willingness to take manual control, increasing enjoyment of automated driving and preventing disuse of automation~\cite{goldman2022trusting}. In a study conducted by Helldin et al., the trust analysis showed that participants who received uncertainty information trusted the system less than those who did not, indicating a more proper trust calibration than in the control group~\cite{helldin2013presenting}. 

\end{description}

In level 4 automation, the vehicle can complete an entire journey without human intervention in specific conditions or environments; however, intervention may still be needed in exceptional situations with sufficient lead time. Certain applications in this category may not necessitate a human driver. At level 5 (full automation), no human intervention is required, and the vehicle can operate under all circumstances~\cite{frison2019resurrecting,hartwich2021improving}. In level 4 and 5 automation, the function of explanations concerns both the internal and external stakeholders. The role of explanations includes: 

\begin{description}

\item[] \textbf{Maintaining situation awareness--} Situation awareness is also relevant in high-level automation, particularly in level 4 automation, where occasional human intervention may be necessary. Providing explanations can facilitate a safe transition from automated to manual driving~\cite{pokam2019principles,kunze2019function}. Research has demonstrated that explanations effectively convey functional and system-wide uncertainties in diverse driving scenarios, such as construction zones, unclear lane markings, heavy traffic, animal crossings in wooded areas, and sudden lane changes~\cite{wang2020watch,du2019look}. Moreover, situation awareness cues can enhance engagement in non-driving related tasks (NDRTs) and increase the enjoyment of automated driving~\cite{fereydooni2022incorporating}. In full autonomy, explanations create a feeling of safety and control, elevating the passenger experience and trust, particularly in complex traffic situations such as intersections, road obstructions, construction sites, pedestrian zones, slow-moving bicycles, and ride delays~\cite{hartwich2021improving,schneider2021explain,eimontaite2020impact,schneider2023don,colley2022effects}.

\item[] \textbf{Increasing driving knowledge--} Commentary driving is a technique that involves drivers verbalizing their observations, assessments, and intentions. Explanations are explored as part of a driving commentary that mimics how a human instructor would explain driving behaviour~\cite{omeiza2022spoken}. It involves aspects such as driving speed, timing of overtaking another vehicle, distance to object, change plan, and lane positioning. By articulating the driving actions out loud, novice drivers can gain a better understanding and awareness of their surroundings~\cite{wayveLINGO1Exploring}. This would also help learners better understand the driving processes and increase their driving knowledge through conversational explanations. Moreover, research has shown that training in commentary driving improves responsiveness to hazards in a driving simulator~\cite{crundall2010commentary}.     

\item[] \textbf{Safe interaction with external stakeholders--} Explanatory signals provided through visual displays play a crucial role in compensating for the absence of direct interaction with human drivers, typically involving gestures and eye contact with other road users. In scenarios such as four-way intersections and pedestrian crosswalks, where communication between AVs and other road users is essential, effective communication mechanisms become even more critical~\cite{avetisyan2023investigating}. One way to improve interaction is by indicating the vehicle's driving mode, whether it is in automated or manual mode. This could foster transparency and help pedestrians understand what to expect from the AV and adjust their actions accordingly~\cite{joisten2020displaying,singer2020displaying}. It would reduce the frustration for internal stakeholders by taking them out of the loop~\cite{wiegand2020d}. External communication is also important in ridesharing scenarios, as it helps users identify the AV and understand its intent, especially when there are multiple AVs on the road~\cite{owensby2018framework}.

\end{description}

Our analysis revealed that the literature predominantly addressed explanations related to level 4 and 5 automation, with limited coverage for levels 1 to 3 across all aspects (See Table \ref{tab:compare}). 


\subsection{RQ2: What are the timings and situations requiring an explanation?} 
\label{sec:situations}
According to our findings, explanations can be classified based on their timing as either proactive or reactive. Proactive explanations are given in anticipation of future needs, while reactive explanations are generated in response to a request after an event has occurred. Proactive explanations are particularly important in low-level automation, where the timing of warnings for potential interventions is a key concern. Studies have shown that providing explanations before the AV takes action promotes more trust than explanations provided after the event~\cite{haspiel2018explanations,koo2015did}. Another aspect to consider is the criticality of the event, as it influences the optimal timing for delivering explanations~\cite {shen2020explain}.
 
In regard to criticality, we have identified two types of situations that require explanations: critical situations that threaten human health and life and non-critical situations that involve harmless but questionable driving behaviour and style.
Adapted from Schneider et al., we discuss the timing and criticality of explanations for four different categories of driving situations: proactive and reactive explanations, which can be generated for critical or non-critical events ~\cite{schneider2021increasing}. It is essential for AVs to be capable of providing all these four categories of explanations in relevant contexts. 


\subsubsection{Proactive explanations in non-critical situations }
\label{sec:pro-non}
This type of explanation aims to increase situation awareness for internal stakeholders by supporting their understanding of ``what is going on''. This includes providing information about the elements within the environment, their meaning, and a projection of their status in the near future~\cite{zang2022effects}. External stakeholders also benefit from the proactive display of the driving mode of the AVs in inconsequential situations~\cite{joisten2020displaying,faas2021pedestrian}. Proactive explanations in non-critical situations are intended to improve the overall user experience and comfort. Nevertheless, further research is needed to identify non-critical situations requiring proactive explanations. Based on the literature, providing stakeholders with proactive explanations in non-critical situations could generate the following benefits:

\begin{description}

\item[] \textbf{Acceptance and trust for new users--} As discussed in Section~\ref{sec:trust}, acceptance and trust rely on an appropriate understanding of the technology. One way to achieve this is by expressing its internal processes proactively. Kuhn et al. and Omeiza et al. presented this type of explanation for normative driving actions, including move, stop, and lane change, e.g., ``Car is stopping because the traffic light is not green on ego’s lane'', ``Stopping because cyclist stopped on my lane''~\cite{kuhn2023textual,omeiza2023effects}. Ruijten et al. added anthropomorphic features to such explanations where AV perform a commentary driving (e.g., ``We’re on a cobbled road with pedestrians, I’m slowing down.'')~\cite{ruijten2018enhancing}. Their findings show that adding a layer of human likeness to an explanatory agent made it perceived to be more intelligent and appealing. Although proactive explanations for non-critical situations help build initial trust and allow constant control over driving behaviour, they quickly become overwhelming and create a high workload for the user~\cite{kunze2019function,fereydooni2022incorporating}. Hartwich et al. suggested adapting this to the user-specific needs to gain broader acceptance rather than providing complete information permanently as a universal standard~\cite{hartwich2021improving}.        

\item[] \textbf{Avert frustration and surprise--} In high-level automation, proactive explanations are necessary for onboard users who do not have any control over driving behaviour. This type of explanation helps to ease frustration and minimize the need for users to constantly ask about the AV's decisions~\cite{gervasio2018explanation}. For instance, in a scenario where the AV stops for longer than usual, it displays a message ``Yielding to pedestrians!" for the riders and gives a view of where the pedestrians are in the scene, what the AV is sensing and how it's reacting~\cite{waymoWaypointOfficial}. 

\end{description}

\subsubsection{Proactive explanations in critical situations}
\label{sec:pro-cri}
Proactive explanations in critical situations play a pivotal role in ensuring the safety of everyone involved, particularly in cases where the driver is required to assume manual control. In such cases, the system should alert the driver regarding potential high-risk situations or the possibility of a failure~\cite{rovira2019looking}. For other road users, proactive communication of the AV's intention is crucial to ensure safe and efficient interaction with them. The purpose of proactive explanations in critical situations is twofold, as discussed below:

\begin{description}
 \item[] \textbf{Alert/Prepare driver to takeover--} Proactive explanations for critical situations require drivers to pay more attention to the road situation and the recommended action to prepare for a potential takeover event~\cite{lee2023investigating}. Chen et al. proposed three types of critical situations that may lead to a takeover event: operational problems, system limitations, and unexpected events~\cite{chen2023adding}. Operational problems are related to the AV's internal functions (i.e., operating system problems involving sensors, computation, and communication between hardware and software) that provide no visible cues within the driving environment. System limitations occur when the AV approaches its operation limits due to various factors such as road or environmental conditions (e.g., construction site, highway exit, traffic jam, foggy weather, road with bends or poor lane markings~\cite{monsaingeon2021indicating}. Unexpected situations include where the AV may be uncertain about its detection or classification of certain road elements and events (e.g., cut-in vehicle or stopped vehicle cause emergency manoeuvres~\cite{kunze2019automation}, failure prediction~\cite{stocco2022thirdeye}, and accident anticipation~\cite{monjurul2021towards}.

 \item[] \textbf{Intent communication for external stakeholders--} Research has extensively examined the external communication of AVs in potentially critical/unsafe situations (e.g., unsignaled intersection~\cite{faas2021pedestrian}, zebra crossing, parking lot~\cite{habibovic2018communicating}, and four street crossing scenarios~\cite{nguyen2019designing} involving pedestrians, cyclists, skateboarders, individuals with disabilities, and other vehicles. Various design elements have been explored in online and simulated environments to facilitate unambiguous communication between AVs and other road users~\cite{verma2019pedestrians,lanzer2023interaction,haimerl2022evaluation,colley2023scalability}. However, it is still unknown how AVs should communicate proactively in critical situations and whether explanations are needed immediately after a critical situation.

\end{description}

\subsubsection{Reactive explanations in non-critical situations}

Reactive explanations are initiated by users in response to unexpected behaviour exhibited by the AV, especially when no apparent cause is evident in the environment. Research has explored various types of questions users may pose, including inquiries about what the system did, why it acted in a certain way, why it did not do something else, and how it arrived at its decision. A study by Graefe et al. found that `why' and `how' questions are preferred for enhancing transparency, understandability, and predictability~\cite{graefe2022human}. Similarly, Wiegand et al. found that users expressed a desire for reactive explanations to influence the AV's driving behaviour interactively~\cite{wiegand2020d}. For instance, users may inform the vehicle that a pedestrian will not cross the street, enabling the AV to proceed safely. Such interaction could potentially correct the AV's driving behaviour. Another research has proposed the use of conversational agents, where AVs provide brief proactive explanations for unexpected driving behaviour, allowing users to inquire for further information about the situation~\cite{schneider2023don}.
 
Several technical-centric studies introduced post-hoc explanation methods to justify a driving action retrospectively in response to a request~\cite{malawade2022roadscene2vec,marchetti2022explainable,charroud2023xdll}. However, much of this work is not comprehensible to non-technical users, which requires them to be integrated into HMIs and further assessed for their usability with human-subject studies.

\subsubsection{Reactive explanations in critical situations}

Reactive explanations for critical situations involve events ranging from near misses to incidents compromising human health and lives. However, existing literature has not sufficiently explored reactive explanations concerning critical situations. In 2020, Wiegand et al. conducted a study on unexpected driving scenarios that do not necessarily result in human injury, hardware damage, or an interruption to the regular operation but cause a great deal of frustration and dissatisfaction for the user~\cite{wiegand2020d}. Some of these scenarios in which the participants requested an explanation include unnecessary lane changes that almost lead to collisions, abrupt stops during turns, unexpected stops, sudden acceleration and deceleration, and unclear interaction with pedestrians.

Moreover, reactive explanations provide evidence for post-incident forensic analysis, as they can answer questions that help determine liability-related issues arising from autonomous decisions~\cite{padovan2023black,rastogi2022explaining}. 
Explainable AI (XAI) techniques are currently capable of providing answers to the following questions: 

\begin{itemize}

\item[-] \textit{What happened?} 
This involves presenting a factual chain of events that demonstrates causation. 
\item[-] \textit{How did it happen?} 
This requires establishing an unbroken chain of causation between the defendant's negligence and the claimant's injury.
\item[-] \textit{Why did it happen? }
This involves identifying the breach of duty.
\end{itemize}

By answering these questions, XAI can help explain the events and the chain of events to establish the factual and legal causation required by common and civil law systems. Consequently, such explanations can fulfil the obligations to establish causation, aiding in identifying the responsible party or the percentile of responsibility.

\begin{table*}[!hbt]
\caption{Influence of task and context on explanatory information concerning internal and external stakeholders}
\label{tab:table1}
\centering
\begin{tabular}{lllll}
\hline
Transparency                            &                   & For internal stakeholders &  & For external stakeholders      \\ \cline{1-1} \cline{3-3} \cline{5-5} 
\multirow{3}{*}{Layer   1: Goal}        & \multirow{3}{*}{} & Driving state             &  & Driving mode                   \\
                                        &                   & Driving action            &  & Intent communication           \\
                                        &                   & Trajectory planning       &  & Information + warning          \\ \hline
\multirow{2}{*}{Layer   2: Reasoning*}   & \multirow{2}{*}{} & Driving action            &  & \multirow{2}{*}{\textit{not addressed}} \\
                                        &                   & Maneuver planning         &  &                                \\ \hline
\multirow{2}{*}{Layer   3: Projections} & \multirow{2}{*}{} & Uncertainty communication &  & \multirow{2}{*}{\textit{not addressed}} \\
                                        &                   & System limitations        &  &                                \\ \hline
\end{tabular}

Adapted from~\cite{chen2018situation}. *Although the literature suggests that agents typically provide explanatory information proactively at any level, there are also instances where reasoning-level explanations are given in response to users' queries.
\end{table*}

\section{Explanatory Information}
\label{sec:info}

Explanatory information enables a clear understanding of the AV's behaviour, facilitating seamless and efficient interaction between humans and automation. Explanatory information, often referred to as `transparency', can take different forms depending on the level of detail required. While certain actions of AVs may be self-explanatory in specific contexts, others may necessitate detailed underlying reasoning to achieve greater transparency. 


\subsection{RQ3: What are the layers of transparency for the explanatory information?} 
\label{sec:RQ3}


In exploring the literature on transparency in autonomous systems, one notable resource is the IEEE standard, which defines transparency requirements at progressive levels~\cite{winfield2022ieee}. At the lower levels, documentation about the automation is provided before any interaction occurs, including the system's general principles of operation, expected behaviour, and sensor data collection and usage. At the higher levels, transparency requires the system to be responsive to user-initiated queries. For our discussion, we focus on transparency at the interaction level, specifically in the context where individuals are directly impacted by the decisions made by an AV.

At the interaction level, the layers of transparency the system requires vary depending on the tasks and purpose. One widely accepted protocol proposed by~\cite{chen2018situation} comprises three layers of transparency. Each layer outlines the necessary information that an agent must communicate to uphold a transparent interaction with the user. The first layer of transparency concerns the goals, actions, and current status of the agent (e.g., ``AV stopping!''). The next layer of transparency contains the reasoning process, beliefs, and motivations towards the current task (e.g., ``AV stopping because the traffic light is red!''). On the last layer of transparency, the agent is expected to predict future outcomes and manage uncertainty and potential limitations (e.g., ``AV stopping because the traffic light is red. It might take longer to take off due to a crowded pedestrian crossing.'').

\subsection{RQ4: How do task and context influence the explanatory information?} 
\label{sec:RQ4}

We found that the task and context influence the explanatory information depending on the level of autonomy and the stakeholder's needs. In light of this, we reviewed the literature to determine the explanatory information necessary for internal and external stakeholders through three layers of transparency (See Table \ref{tab:table1}).

\subsubsection{Layer 1. Transparency to provide current state, goal, and action}
\label{sec:layer1}
This layer of transparency entails the display of basic driving status cues to internal stakeholders, such as estimated arrival time, navigation, and traffic conditions~\cite{fereydooni2022incorporating,hartwich2021improving}. In collaborative driving settings, the driver must be aware of the current mode and status to avoid confusion. During autonomous driving, the driver must be informed that the vehicle is following established driving norms and traffic regulations, such as indicating the current speed and speed limit. This layer of transparency needs no descriptive explanations; it only sends proactive, non-critical short messages to support situation awareness, e.g., ``Construction site!''~\cite{schneider2021increasing}. Moreover, the driver can detect and understand the actions being performed by the vehicle through the visual arrows/carpet shown in HMI displays, indicating the direction and route without providing detailed reasoning~\cite{pokam2019principles,zhang2023tactical}. Concerning the planning task, several approaches have been suggested, such as bird’s eye view displays for scene comprehension to identify the most influential objects in planning a trajectory~\cite{limeros2022towards,carrasco2021scout,brewitt2021grit,renz2022plant,zhang2022trajectory}.

Regarding external stakeholders, as per the Turing Red Flag law, AVs must be designed to clearly identify themselves as such and indicate their presence when interacting with external agents~\cite{winfield2022ieee}. Research has explored the effect of external human-machine interfaces (eHMIs) to display the vehicle's driving mode~\cite{joisten2020displaying,faas2020external,singer2020displaying}. Furthermore, literature has investigated the use of eHMIs in ambiguous crossing and unsignaled intersections for intent communication to facilitate safety crossing~\cite{colley2022effects,lanzer2023interaction,colley2022investigating}.

\subsubsection{Layer 2. Transparency to provide reasoning behind AV’s actions and decisions}
\label{sec:layer2}
In this layer of transparency, it is essential for users to comprehend the reasoning behind a vehicle's actions. This involves understanding what the AV is perceiving, analysing, and making decisions. The user could trigger this function to produce a brief and immediate explanation of why the system acted in a certain way in a given situation~\cite{schneider2023don,graefe2022human}. Alternatively, the system itself could initiate a brief explanation of its ongoing activities~\cite{du2019look,zhang2023tactical,graefe2022human}. Explanations generated for driving action could have different levels of specificity (abstract or specific) depending on the user's desire to influence the driving activity~\cite{omeiza2023effects}. Furthermore, such explanations can facilitate interactive exploration of hypothetical situations, enabling users to inspect `what if' scenarios.

Research has yet to delve into the communication of reasoning-level explanations and beyond with external stakeholders, particularly in the vehicle-to-everything (V2X) context. Such information is difficult to direct to individuals; it is, per the definition, broadcasted~\cite{dey2020taming}. AV reasoning is often conveyed through its driving behaviour, which constitutes implicit communication. Further research is needed to identify scenarios necessitating explicit communication for AV reasoning. Additionally, there is a concern regarding handling communication with the outside in case eHMIs malfunction or display misleading information~\cite{hollander2019overtrust}. Given the limitations of LED-based eHMIs, alternative methods, such as personalized messages through smart and wearable devices, may be more appropriate for conveying this layer of transparency to other road users.  

\subsubsection{Layer 3. Transparency to convey uncertainty and predictions of failure}

In this layer of transparency, the driver must know the AV's capabilities and the conditions under which it operates. The driver should be informed of any unexpected outcomes, uncertainties, or limitations when the AV approaches its operational boundaries due to road or environmental conditions, such as roads with curves, unclear markings, or heavy traffic~\cite{chen2023adding}. This can be achieved through a feedback mechanism that indicates an AV's proximity to its limits and control transition~\cite{monsaingeon2021indicating}. Studies have shown that informing drivers of the AV's uncertainty can better prepare them to switch to manual control when required~\cite{helldin2013presenting}.

In 2022, Shull et al. implemented a system that offers preemptive feedback to the user when the AV fails to detect lane lines, thereby alerting them before it veers out of its designated lane~\cite{shull2022using}. Likewise, Kruger et al. and Kunze et al. conducted research on the impact of conveying reduced sensor reliability and machine certainty in diverse weather conditions, such as fog and rain~\cite{kruger2020feeling,kunze2019conveying}. 

Research has shown that communicating potential uncertainties can enhance trust calibration and situation awareness, leading to safer takeovers~\cite{kunze2019automation}. Issuing a warning in such circumstances is crucial, although the level of detail required may vary depending on the level of situational awareness required. In some cases, presenting visual information alone may be sufficient~\cite{colley2021investigating}.

\section{Explanatory Information Communication}
\label{sec:com}

Human-machine interfaces (HMIs) are used to communicate information from AVs to onboard and surrounding road users. Bengler et al. and Murali et al. provided a taxonomy for different types of HMIs on AVs~\cite{bengler2020hmi,murali2022intelligent}. 
HMIs are classified as `internal' and `external', comprising all the interfaces within the vehicle's interior and exterior. 
We identified three key design considerations for effective communication: conspicuousness (easy to perceive), comprehensibility (easy to understand), and interactivity (easy to engage with). Table \ref{tab:communication} summarises the literature through the lenses of these design considerations. The review results show that, regarding internal HMIs, there has been increasing attention on head-up displays (HUDs) that incorporate different visual elements to enhance conspicuousness and comprehensibility. As for external HMIs, the LED display has been the primary focus, featuring instructional and symbolic elements. Although interactivity is partially addressed in relation to internal HMIs, further exploration is needed for both internal and external stakeholders concerning bidirectional communication where users actively engage with the system and provide feedback.

\begin{table*}[!hbt]
\centering
\caption{Qualitative assessment of the reviewed papers in terms of communication methods for internal and external HMIs}\label{tab:communication}
\begin{adjustbox}{width=1\textwidth}
\begin{tabular}{llllclclc}
\hline
                                                    &                             & \textbf{Communication method}         &  & \textbf{Conspicuousness} &  & \textbf{Comprehensibility} &  & \textbf{Interactivity} \\ \cline{1-3} \cline{5-5} \cline{7-7} \cline{9-9} 
\multicolumn{1}{l|}{\multirow{16}{*}{{\rotatebox[origin=c]{90}{\textbf{Internal HMI}}}}} & \multirow{7}{*}{\rotatebox[origin=c]{90}{Medium}}     & Head-down displays   (HDDs)  &  & \ding{119}          &  & \ding{119}            &  & \ding{109}          \\ \cline{3-9} 
\multicolumn{1}{l|}{}                               &                             & Head-up displays   (HUDs)    &  & \ding{108}           &  & \ding{108}             &  & \ding{119}        \\ \cline{3-9} 
\multicolumn{1}{l|}{}                               &                             & Head-mounted displays (HMDs) &  & \ding{109}            &  & \ding{109}              &  & \o{}           \\ \cline{3-9} 
\multicolumn{1}{l|}{}                               &                             & Light band                   &  & \ding{109}            &  & \ding{109}              &  & \o{}           \\ \cline{3-9} 
\multicolumn{1}{l|}{}                               &                             & Vibrotactile feedback        &  & \ding{109}            &  & \ding{109}              &  & \o{}           \\ \cline{3-9} 
\multicolumn{1}{l|}{}                               &                             & Audio voice message          &  & \ding{119}            &  & \ding{119}              &  & \ding{119}          \\ \cline{3-9} 
\multicolumn{1}{l|}{}                               &                             & Sonification (music)         &  & \ding{109}            &  & \ding{109}              &  & \o{}           \\ \cline{3-9} 
\multicolumn{1}{l|}{}                               &                             &                              &  &                 &  &                   &  &               \\ \cline{3-9} 
\multicolumn{1}{l|}{}                               & \multirow{5}{*}{\rotatebox[origin=c]{90}{Features}}   & Symbolic                     &  & \ding{119}          &  & \ding{119}            &  & \ding{109}          \\ \cline{3-9} 
\multicolumn{1}{l|}{}                               &                             & Informative (text)           &  & \ding{119}            &  & \ding{119}              &  & \ding{109}          \\ \cline{3-9} 
\multicolumn{1}{l|}{}                               &                             & Image (instrument cluster)   &  & \ding{119}          &  & \ding{119}            &  & \o{}           \\ \cline{3-9} 
\multicolumn{1}{l|}{}                               &                             & Graphs/plots                 &  & \ding{119}            &  & \ding{119}              &  & \o{}           \\ \cline{3-9} 
\multicolumn{1}{l|}{}                               &                             & Bird’s eye view              &  & \ding{109}            &  & \ding{109}              &  & \o{}           \\ \cline{3-9} 
\multicolumn{1}{l|}{}                               &                             &                              &  &                 &  &                   &  &               \\ \cline{3-9} 
\multicolumn{1}{l|}{}                               & \multirow{2}{*}{\rotatebox[origin=c]{90}{Mult.}} & Audiovisual            &  & \ding{119}          &  & \ding{119}            &  & \o{}           \\ \cline{3-9} 
\multicolumn{1}{l|}{}                               &                             & LED and vibrotactile         &  & \ding{109}            &  & \ding{109}              &  & \o{}           \\ \cline{3-9} 
                                                    &                             &                              &  &                 &  &                   &  &               \\ \cline{3-9} 
\multicolumn{1}{l|}{\multirow{11}{*}{{\rotatebox[origin=c]{90}{\textbf{External HMI}}}}}  & \multirow{3}{*}{\rotatebox[origin=c]{90}{Medium}}     & LED band/strip               &  & \ding{108}          &  & \ding{108}            &  & \o{}           \\ \cline{3-9} 
\multicolumn{1}{l|}{}                               &                             & Displays                     &  & \ding{119}          &  & \ding{119}              &  & \ding{109}          \\ \cline{3-9} 
\multicolumn{1}{l|}{}                               &                             & Projection                   &  & \ding{119}            &  & \ding{119}              &  & \o{}           \\ \cline{3-9} 
\multicolumn{1}{l|}{}                               &                             &                              &  &                 &  &                   &  &               \\ \cline{3-9} 
\multicolumn{1}{l|}{}                               & \multirow{4}{*}{\rotatebox[origin=c]{90}{Features}}   & Instructional                &  & \ding{119}            &  & \ding{119}              &  & \ding{109}          \\ \cline{3-9} 
\multicolumn{1}{l|}{}                               &                             & Symbolic                     &  & \ding{119}          &  & \ding{119}            &  & \o{}           \\ \cline{3-9} 
\multicolumn{1}{l|}{}                               &                             & Metaphorical                 &  & \ding{109}            &  & \ding{109}              &  & \o{}           \\ \cline{3-9} 
\multicolumn{1}{l|}{}                               &                             & Anthropomorphic              &  & \ding{109}            &  & \ding{109}              &  & \ding{109}          \\ \cline{3-9} 
\multicolumn{1}{l|}{}                               &                             &                              &  &                 &  &                   &  &               \\ \cline{3-9} 
\multicolumn{1}{l|}{}                               & \multirow{2}{*}{\rotatebox[origin=c]{90}{Mult.}} & Display and projection       &  & \ding{109}            &  & \ding{109}              &  & \o{}           \\ \cline{3-9} 
\multicolumn{1}{l|}{}                               &                             & Audiovisual                 &  & \ding{109}            &  & \ding{109}              &  & \o{}           \\ \hline
\end{tabular}
\end{adjustbox}
\\[5pt]
\ding{108} addressed, \ding{119} partially addressed, \ding{109} barely addressed, \o{} not addressed, Mult., Multimodality  
\end{table*}

\subsection{RQ5: How is the explanatory information communicated to internal stakeholders?} 
\label{sec:RQ5}
The internal HMIs, encompassing all interfaces within the vehicle interior, allow explanatory information to be conveyed to onboard users. It communicates the system's status, intention, and motion decisions transparently, ensuring mode awareness and a smooth transition between driving modes for a comfortable and safe journey. The internal HMI modalities are categorized into visual, audio, and vibrotactile feedback.

\subsubsection{Visual feedback}
In the realm of in-vehicle visualizations, feedback devices such as head-down displays (HDDs), head-up displays (HUDs), head-mounted displays (HMDs), and light bands/strips are commonly studied. While HDDs offer the benefit of not obstructing the users' view of the outside world, they can still be a source of distraction as users may become too engrossed in the display instead of the road ahead~\cite{morra2019building}. On the other hand, HUDs are designed to minimize driver distraction by presenting essential information relative to the positional and temporal environment within the driver's primary field of vision. Research suggests that HUDs offer better driving experiences over HDDs, resulting in decreased cognitive load, higher usability, and better performance control~\cite{colley2021investigating,smith2016head,kunze2019automation}. Augmented Reality HUDs (AR-HUDs) as Windshield Displays (WSDs) are the next step in developing HUDs by covering the entire windshield. These innovative tools have demonstrated a remarkable capacity to enhance a driver's intuitive cognition and foster an efficient driving experience, particularly in challenging driving scenarios~\cite{morra2019building,colley2021investigating}. However, to realize these benefits, HUD information must be thoughtfully designed in terms of content, timing, and placement, ensuring it complements rather than conflicts with real-world information.

Head-mounted displays (HMDs) offer the same advantages as HUDs. Optical HMDs allow users to access necessary information while remaining aware of their surroundings. Discrete HMDs provide a fully immersive VR experience, which can replace the physical world with a virtual environment. This has the potential to reduce distractions and increase productivity for riders. Fereydooni et al. suggested incorporating short visual cues in the VR world during a ride, which was found to be helpful by passengers~\cite{fereydooni2022incorporating}. 
However, these devices are limited in the market and require further research to determine their suitability for in-vehicle applications.

Various visual presentation strategies were employed to increase the conspicuousness and comprehensibility. Kunze et al. suggested using manipulable, abstract signs to communicate uncertainties, with visual variables adjusting in size and transparency based on the level of ambiguity~\cite{kunze2018augmented,kunze2019function}. Zhang et al. proposed using AR carpet visualizations to display lane availability, with arrows indicating lane-changing suggestions~\cite{zhang2023tactical}. Monsaingeon et al. implemented an HDD interface that contains the instrument cluster, distance to the followed vehicle, detected road markings, system activation, and a small area for textual messages~\cite{monsaingeon2021indicating}. Colley et al. presented semantic segmentation visualizations of the most relevant objects (e.g., vehicles, pedestrians) projected on the windshield as AR-HUD to promote user trust and situational awareness~\cite{colley2022effects}.

Given the technical difficulties and significant construction challenges in realising AR-HUDs and HMDs, current displays often come in the form of prototypes, concepts or demonstration videos. While AR-HUD proposes a more granular possibility of highlighting all relevant traffic objects, one of the concerns is that visual clutter may cause driver distraction and negatively impact driving performance. Kunze et al. argued that using the instrument cluster to visualise uncertainty information can increase mental workload~\cite{kunze2019conveying}. Therefore, less distracting visualisation methods such as light bands (light strips) are used as a peripheral cue to bring attention to the objects of importance in the scene. Some works proposed lighting certain parts of the band depending on the position of the crossing pedestrian to indicate intention and perception~\cite{colley2021should,wilbrink2020reflecting,hecht2022users}.

\subsubsection{Audio feedback}

Auditory feedback serves as another effective communication channel with internal stakeholders. Researchers have explored different strategies, such as speech, blended sonification, and alerting sounds. In a study conducted by Avetisian et al., audio messages are used to direct users' attention to critical objects in the traffic to indicate their significance for the AV's decisions~\cite{avetisian2022anticipated}. Eimontaite et al.'s study revealed that sound-based feedback improved the ease of operation and journey experience for elderly users when examining the impact of different modalities~\cite{eimontaite2020impact}. Zhang et al.'s research demonstrated the differences in the effectiveness of audio explanations across different age groups~\cite{zhang2021drivers}. They discovered that voice-based explanations provided before action was taken resulted in the highest trust among elderly drivers. Additionally, blended sonification, which manipulates background music, was proposed to convey the automation's reliability level to keep drivers informed. This strategy effectively increased monitoring behaviour and reduced response time to takeover requests, as suggested in studies by~\cite{chen2023adding,chen2021manipulating}.

\subsubsection{Vibrotactile feedback}
Vibrotactile feedback is explored as an unobtrusive and intuitive means of communicating uncertainty to users. In 2020, Kruger et al. conducted a study on the effects of using a vibrotactile belt to communicate sensor uncertainty information~\cite{kruger2020feeling}. Their results showed that participants were able to understand the encoded information and perceived it as a meaningful communication medium. However, further investigation is needed to determine the usability of vibrotactile feedback for the ageing population, as it is known that age is associated with sensory and cognitive decline. In 2019, Kunze et al. suggested using vibration motors mounted in the vehicle seat to convey uncertainty~\cite{kunze2019conveying}. While vibrotactile feedback alone can be helpful, it can also cause sudden spikes in attention, making it less effective for conveying dynamic information. Their work suggests that combining vibrotactile feedback with other modalities may be more effective in communicating such information than relying solely on it.

\subsubsection{Multimodality for internal stakeholders} 
\label{sec:multion}
Multiple feedback modalities enable the perception of a large amount of information without overwhelming a single sensory input. Multimodality is also necessary to convey warnings in case human intervention is needed. This can increase the sense of urgency and lead to quicker response times~\cite{bengler2020hmi}. However, it is essential to consider the form and content of multimodality depending on the context. These modalities should complement each other and aim to reduce information overload and clutter. Research has shown that a combination of visual and auditory resources is an effective feedback strategy to increase situational awareness~\cite{avetisian2022anticipated,colley2023scalability,zhang2022trajectory}. In 2022, Shull et al. studied how uncertainty feedback, representing current capability, influenced the transition of control to manual driving~\cite{shull2022using}. Their results showed that providing auditory-visual feedback led to longer HMI viewing times and early takeovers. In contrast, Zhang et al. found that auditory-visual explanations decreased decision-making performance~\cite{zhang2023keeping}. Alternatively, visual-haptic feedback is introduced to raise the driver's attention when distracted by other tasks. As proposed in Kunze et al.'s study, combining vibrotactile feedback with peripheral light can enhance the effectiveness of the feedback, as both can be adjusted based on the levels of uncertainty~\cite{kunze2019conveying}. Nevertheless, further research is needed to assess the practicality of this approach.

\subsection{RQ6: How is the explanatory information communicated to external stakeholders?} 
\label{sec:RQ6}
External stakeholders perceive explanatory information through external HMIs, which are installed on or projected from the surface of a vehicle. Although standardized eHMIs like indicators and brake lights are already in use, new forms of eHMIs are being explored to enable seamless communication with other road users. While forward-facing displays are the primary focus of research, various other visual concepts, including optical projections and LED bands, are also being studied to convey information about the AV's behaviour and status. Overall, the literature suggests that all proposed communication concepts were better received than having no external communication at all. Nevertheless, these modalities may be less effective for distracted pedestrians who may not be able to correctly perceive, understand, and interpret the information.

\subsubsection{Forward-facing displays}
\label{sec:forw}
Various concepts have been suggested for positioning forward-facing displays on the vehicle's surface, including the bumper, grille, hood, windows, and windshield. This kind of display signals information in four ways: instructional, symbolic, anthropomorphic, and metaphorical~\cite{verma2019pedestrians}.   

Instructional signals are usually presented as text, providing advisory and informative messages. Advisory signals are not favoured for safety reasons (e.g., AV shows ``Cross", but another vehicle in the adjacent lane may not stop) \cite{habibovic2018communicating}. Instead, intention-based informative messages (e.g., ``Stopping") are recommended by ISO~\cite{ISO23049:2018}. Colley et al. considered instructional signals as the bidirectional communication between the AV and the pedestrian~\cite{colley2021investigating}. The study found that participants appreciate it when an AV responds to their hand gesture of ``Thank you!" with the feedback ``You are Welcome!". They described it as more human-like, friendly, and clear. As useful as it is, there could be other limitations to textual signals, including language barrier, illegibility due to distance, and vision disability. 

The symbolic visual signals feature patterns (e.g., traffic symbols) that would animate over the LED display. Singer et al. explored the impact of animated symbols, like the P sign, arrow, and flashing hand, on driving behaviour and intent recognition~\cite{singer2020displaying}. The results show that a combination of signals helps a better perception of the AV's intention and increases the perceived safety. Avetisyan et al. focused on eHMI for communication with human drivers in conventional vehicles in mixed traffic situations~\cite{avetisyan2023investigating}. The visual messages displayed on the front of the AV warned drivers of uncertain situations, leading to increased awareness and trust. 

The anthropomorphic signals comprised facial expressions such as animated eyes that would blink or gaze in different directions to indicate the vehicle's intention~\cite{rouchitsas2022ghost}. In 2023, Colley et al. experimented with an expressive eHMI for autonomous vehicles~\cite{colley2023scalability}. They utilized a bumper display shaped like a mouth, which would remain neutral with a horizontal line when the AV was not yielding and turn into a smile when the AV was yielding to pedestrians. 
Finally, metaphorical pictographs use animated allegorical narratives (e.g., `walking man' at pedestrian crossings) to convey pedestrians to exercise caution, wait, or cross the street. Metaphorical pictographs and anthropomorphic signals seem most easily understood, especially as the characters explicitly display gestures and movements to indicate the future steps pedestrians could follow~\cite{verma2019pedestrians}.

\subsubsection{Optical projections}

An alternative means of conveying information is projecting messages---trajectories, stopping points, intentions, and directions---on the road around the vehicle. This medium allows for the highest amount of information to be communicated, as the interface can be designed in detail~\cite{dey2020taming,bengler2020hmi}. In an earlier study, Nguyen et al. proposed visualisations displayed on-road projections to indicate the intention of AV, such as stopping, slowing down, or proceeding~\cite{nguyen2019designing}. Although this information helped participants adjust their actions, they demanded a display of timing during the crossing to know when the AV would start moving. 
In 2021, Colley et al. compared the effectiveness of on-road projections versus information displayed on a car's surfaces, such as the bumper and windows~\cite{colley2021investigating}. The results showed that projecting the information on the sidewalk did not lead to increased trust. However, this concept was better at increasing situational awareness and warning distracted pedestrians about oncoming traffic. In another study by Colley et al., street projection was among the best-performing concepts as it was the only display communicating where the AV will stop~\cite{colley2023scalability}.
Despite its potential benefits, on-road projections have certain limitations. They require near-perfect environmental conditions, including lighting, weather, and road surface, to work properly. A powerful projector would also be necessary since the eHMI is projected over a significant distance.

\subsubsection{LED bands}

Light-band exudes light patterns, such as running from one side to another or flashing to convey a message. The advantage of light band eHMI is its technical feasibility compared to projections or forward-facing displays. Furthermore, light bands are visible to multiple pedestrians and are not subject to the constraints of textual messages. Various lighting colours and patterns have been suggested to describe vehicle mode and intent~\cite{habibovic2018communicating}. One idea proposed by Lanzer et al. is to use a light strip that displays the vehicle's intent using different light patterns, such as yellow dots or flashed turquoise, to communicate whether the car is moving or intends to stop~\cite{lanzer2023interaction}. In 2021, Faas et al. implemented a slowly flashing blue-green light above the windshield to signal yielding intent, which helped pedestrians calibrate their trust accordingly~\cite{faas2021pedestrian}. Similarly, in 2023, Zhanguzhinova et al. used red and green lights that switched between moving and stopping actions, resulting in smooth crossing behaviour and increased trust~\cite{zhanguzhinova2023communication}. 


\subsubsection{Multimodality for external stakeholders}

In the realm of eHMIs, while the literature provides instances of combining two or more modalities, the exploration of multimodality has generally been less extensive compared to single-modality approaches~\cite{colley2022investigating,dou2021evaluation}. One line of research explored multimodality to address the inadequacy of current concepts for stakeholders with varying disabilities~\cite{haimerl2022evaluation}. They compared auditory-visual modalities for individuals with and without intellectual disability (ID) to assess the inclusiveness of current eHMI concepts. The study involved emitting an audio message from the front of the vehicle for pedestrians and displaying a visual signal on the hood to indicate yielding intention. This implementation of multimodal eHMIs has shown positive results in terms of ID inclusivity. Regardless, it is essential to consider other factors, such as visual impairment, distracted road users, and language barriers, as relying solely on visual displays may hinder safe interaction with various road users. There is no ``best" modality, as acceptability depends on various factors and is not universal, which calls for further research in mixed-modality eHMIs.

\begin{figure*}[htb!]
\centering
  \includegraphics[width=\textwidth]{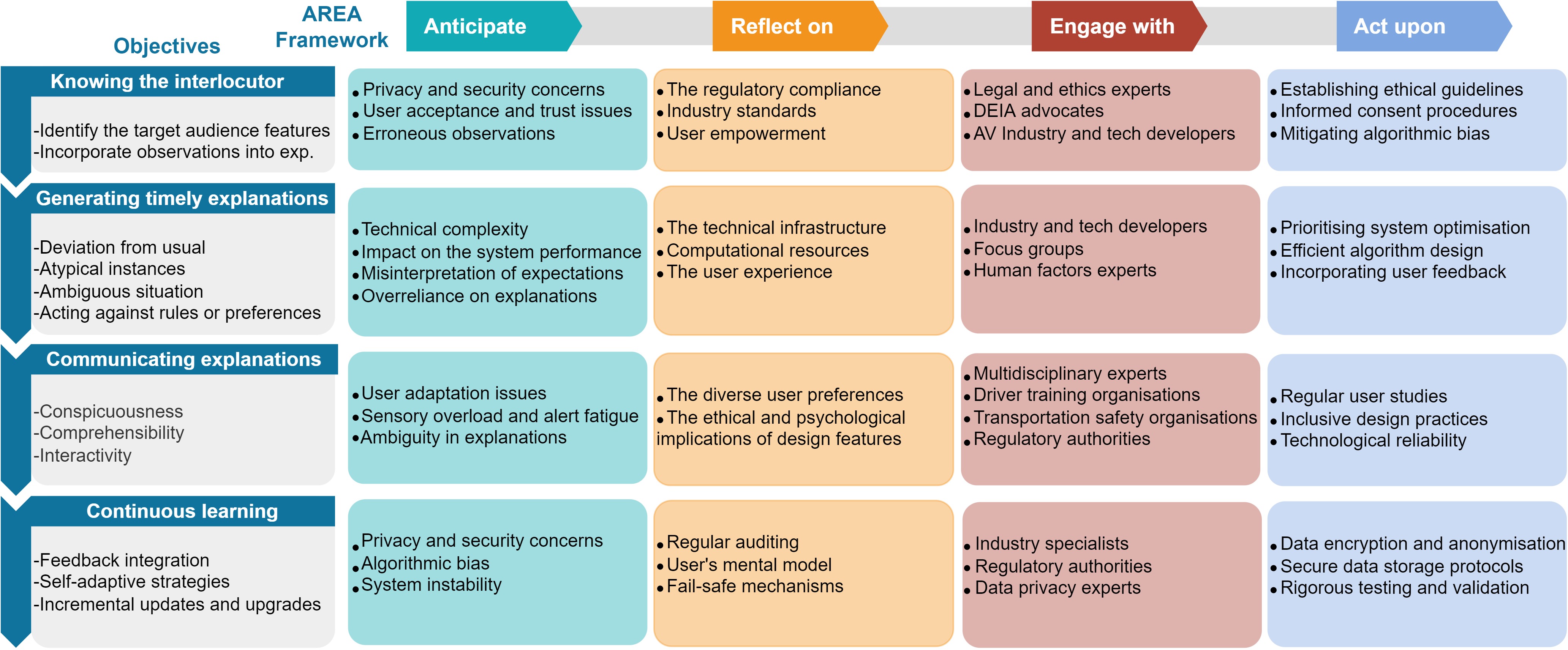}
 \caption{Responsible research and innovation grounded roadmap for advancing explainable AV systems}
 \label{fig:roadmap}
\end{figure*}

\section{A Responsible Research and Innovation (RRI) Grounded Roadmap}
\label{sec:road}

As shown in our review (Section~\ref{sec:task}-\ref{sec:com}), the explainability of AVs has been widely addressed in the research community. However, there is currently a lack of a systematic approach to the design of explanations, which may lead to suboptimal design of explanatory information or even contradictory outcomes. To address this gap, we suggest a roadmap that designers of explanatory information in the context of AV could utilise to ensure that all key design elements are captured. The roadmap is grounded in responsible research and innovation principles outlined in the AREA framework (See Section \ref{sec:meth}). It pinpoints key design elements related to explanatory task, information, and communications, emphasizing the importance of understanding the interlocutor, generating timely explanations and communicating human-friendly explanations. In contrast to previous discussions (e.g., \cite{omeiza2021explanations,atakishiyev2021explainable,xing2021toward}), our roadmap also integrates elements of continuous learning and improvement. 


\subsection{Knowing the interlocutor}
An interlocutor in this context refers to the Class A users who are directly influenced by the actions taken by an AV (See Section~\ref{sec:user}). Previous research has discussed the integration of driver's physical and cognitive state information into in-vehicle interaction applications~\cite{murali2022intelligent,jansen2022design}. However, these observations have not been incorporated into explanations, as noted in Section \ref{sec:task}. This highlights the need for vehicle interaction technologies to identify the target interlocutor, i.e., the audience, whether the internal or external stakeholder and observe and recognise their physical and cognitive state to generate timely and meaningful explanations. As shown in Section~\ref{sec:situations}, there are several scenarios where incorporating observations into explanations might be necessary, such as:  

\begin{itemize}

\item[-] The observations of stakeholders can be incorporated into explanations when the automation raises an intervention due to detecting deficient states of the driver (e.g., AV take over control due to detecting alcohol breath or the driver's drowsiness).

\item[-] Observation can help offer proactive explanations in situations where there is a risk that the passenger(s) might feel discomfort, confusion, or frustration by a driving-related situation, especially when there is no visible reason exist in the scene.

\item[-] Observations can help to justify an action that resulted from detecting an unusual situation (e.g., AV pulls over and contacts an operator due to detecting the passenger's medical emergency).

\item[-] Understanding external stakeholders' attributes is essential to provide explainable trajectory planning for onboard users, especially when other road users' intentions are not clear but could influence the behaviour of the AV.

\item[-] Detecting relevant characteristics of safety-critical road users (e.g., child, elderly, pet/animal, distracted pedestrians, e-scooter user), if there exists any impairment (e.g., cognitive, vision, hearing, mobility) is necessary to properly calibrate the external communication mode and efficiently interact with them.

\end{itemize}

While an explainable AV with these capabilities can significantly improve user experience and safety, it also poses certain risks and implications related to the effectiveness and ethical use of the data. The literature evaluated several risks and implications concerning such technology, which relates to the explanatory task as follows:

\begin{description}

\item[] \textbf{Data privacy and security concerns--} Incorporating stakeholder observations for explanations raises privacy concerns, especially when sensitive information about drivers or passengers is involved. Proper data handling and anonymization protocols must be in place, and clear guidelines should be established for collecting, storing, and processing stakeholder observations to protect the privacy and confidentiality of the individuals involved~\cite{krontiris2020autonomous,murali2022intelligent}. 

\item[] \textbf{User acceptance and trust--} Users may have concerns about the collection and use of personal data~\cite{murali2022intelligent,jansen2022design}. Therefore, it is essential to empower users to have control over their data and the ability to select which part of observations to opt in or out of the system. It is important to make sure that users are informed about the benefits and implications of incorporating observations in enhancing the AV system's operations.

\item[] \textbf{Erroneous observations--} There is a risk of misinterpreting stakeholder behaviour, leading to incorrect or inappropriate responses. To minimize this risk, it is necessary to have robust AI models capable of accurately understanding and responding to complex human behaviours and intentions~\cite{ma2022survey}. Implementing robust sensor technologies and data fusion techniques can help ensure the accuracy and reliability of stakeholder observations, reducing the risk of false positives.

\item[] \textbf{Legal and regulatory compliance--} Incorporating stakeholder observations for decision-making raises questions about legal liability and compliance with existing regulations. The legal implications of AV interventions and data usage must be thoroughly evaluated and addressed to ensure adherence to relevant laws and regulations~\cite{bengler2020hmi}. Additionally, clear ethical guidelines and informed consent procedures should be established for the collection and use of stakeholder observations in AV decision-making processes.

\item[] \textbf{Algorithmic bias and fairness-- }Integrating stakeholder observations may introduce biases in the AV's decision-making processes, particularly if the observations are not adequately representative or if the algorithms are not properly calibrated. Implementing bias detection mechanisms, algorithmic fairness assessments, and regular audits can help identify and mitigate potential biases, ensuring fair decision-making by the AV system~\cite{gyevnar2022cars,krontiris2020autonomous}.

\end{description}

Addressing these risks and implications requires a collaborative and inclusive approach that combines robust technical solutions, ethical guidelines, and legal compliance. Researchers have been engaging with various relevant parties to enable the responsible and ethical deployment of AV technology through industrial collaborations and workshops~\cite{locken2022accessible,nannini2023explainability,graefe2022human}. We identified some key interest groups to involve in the process, including:

\begin{description}
    
\item[] \textbf{Legal authorities and ethics committees--} Collaborating with legal experts can help establish ethical guidelines, informed consent frameworks, and regulatory compliance measures for integrating stakeholder observations in AV decision-making processes.

\item[] \textbf{DEIA (Diversity, Equity, Inclusion, and Accessibility) advocates--} Involving accessibility, diversity and inclusion advocates can help address the diverse needs of users. Discussions around the importance of fair representation, assessments, and bias detection mechanisms will ensure that the AV system's decision-making processes are inclusive and fair~\cite{patel2023robotics}.

\item[] \textbf{AV industry and technology developers--} Collaborating with the AV industry and technology developers can provide insights into best practices, such as sensor technologies and data fusion techniques, for integrating stakeholder observations in AV systems. This knowledge exchange would also foster consensus on technical specifications and protocols, leading to standardized frameworks.

\end{description}

\subsection{Generating timely explanations}

Providing explanations for every action, even in routine situations, may result in information overload for users, ultimately diminishing the value of explanations and reducing their effectiveness~\cite{ekman2017creating}. Prioritizing event criticality for explanations while ensuring a balance between transparency and simplicity is essential to avoid overwhelming users with unnecessary details. While this has been partially addressed in the literature, further research is needed to focus on timing for proactive explanations (See Section~\ref{sec:pro-non} and \ref{sec:pro-cri}). Timely explanations should address discrepancies between what the user expects the system to do and what it does, thereby promoting a seamless experience. Motivated by Gervasia et al.,~\cite{gervasio2018explanation} work, we suggest a set of guidelines for generating timely explanations that address expectation violation as follows:  

\begin{description}

\item[] \textbf{Deviation from usual--} AV should identify if an action differs from past behaviour in similar situations through statistical analysis of performance logs and semantic models. The explanation should acknowledge the atypical action and explain the reason when it exhibits unusual behaviour (e.g., AV not yielding to pedestrians).

\item[] \textbf{Atypical instances--} AV should generate explanations for actions taken in unusual situations causing atypical action. The AV can identify these situations by their frequency of occurrence and describe the unusual situation to the user (e.g., sharp brake due to an unidentified object—a deer jumping on the road).

\item[] \textbf{Ambiguous situation--} In some situations, the apparent cause of the action may be unavailable to the user, and the user might assume that the information used for the decision comes only from the vehicle's vision sensors. AV should explain the potential mismatch and indicate decision criteria in such situations, e.g., in a ride-sharing scenario, AV divert to an unusual route, justifying that it will take a new rider on the way.

\item[] \textbf{Acting against rules or preferences--} In typical driving conditions, AVs seek to satisfy the established traffic rules, regulations, and preferences; nevertheless, various factors (e.g., to prevent a potentially more severe incident, time frame, physical restrictions) may cause a violation, leading to the agent operating contrary to its set-up. Explanation, in this case, involves acknowledging the violated directive or preference and providing why that occurred (e.g., changing the preferred driving style or stopping at the ``No stopping" sign).

\end{description}

Implementing an explainable AV system with the outlined capabilities can significantly improve user trust and experience. However, it is crucial to consider and act on several key aspects to ensure the effectiveness of the explanations. One such aspect is technical complexity and system performance. Generating context-sensitive explanations requires a robust technical infrastructure and algorithms without impacting the overall system performance and responsiveness~\cite{zablocki2022explainability}. Therefore, it is crucial to prioritize system optimization and efficient algorithm design to ensure that the explanation generation process does not compromise the overall performance and computational resources of the AV system. Another aspect to consider is the risk of misinterpreting user expectations, leading to misaligned explanations that fail to address the user's concerns adequately. To mitigate this risk, incorporating robust user feedback mechanisms and conducting regular user studies can help align the AV system with user expectations more effectively~\cite{schneider2023don}. Additionally, there is a risk of users becoming overly dependent on the explanations, which may lead to a lack of critical thinking and reduced user engagement with the driving task~\cite{zang2022effects}. 

Academics and industrial partners have been working together to address these concerns and fuse their expertise to design AV systems that are both robust and reliable~\cite{novakazi2022design,wang2020watch}. In addition, collaborating with human factor experts in the field can offer valuable insight into users' cognitive processes, behaviours, and preferences~\cite{dey2020taming}. Such collaborations can inform the explanation generation process and user-centred design principles to ensure the successful deployment of explainable AV technology with these capabilities.

\subsection{Communicating human-friendly explanations}

In the literature, there is a consensus that explanations should convey relevant information in a clear, concise, and friendly manner, revealing the reasoning behind a decision, the current state of affairs and the potential outcomes in the future~\cite{omeiza2022spoken,kruger2020feeling,candela2023risk}. For external stakeholders, communication should compensate for the absence of human-driver interaction. While discussions on these aspects exist, the lack of a comprehensive set of guidelines remains a gap. Our approach introduces three key properties for communicating human-friendly explanations in the context of AVs: conspicuousness, comprehensibility, and interactivity. We proposed a thorough approach with a detailed map outlining how to take on each one properly. 

 
\subsubsection{Conspicuousness}
This refers to the degree to which the resulting explanation is clear and easy to notice through the visual, auditory, or tactile channel. The communication channel should be adjusted based on the degree of priority of the message.

\begin{description}
    
\item[] \textbf{Low-priority messages--} These refer to the messages conveyed in layer 1 transparency that describe the current state of the system when it is running smoothly (See Section~\ref{sec:layer1}). To avoid disrupting users while they perform NDRTs, one could consider presenting low-priority messages as visual feedback incorporated into AR windshield displays~\cite{riegler2019augmented}. Visual feedback may include information about other vehicles in close proximity to the ego vehicle, such as their orientation, size, movement, and intention~\cite{wintersberger2020explainable}. This may also include information about detecting partially obstructed objects in the environment, as well as large objects that may block the view of potential hazards~\cite{suchan2021commonsense}. As discussed in Section~\ref{sec:pro-non}, this information should be readily available, especially during the introductory phase, to enhance the perception of safety. Concerning external road users, the low-priority messages involve the display of driving mode when the vehicle is operated in autonomous driving (See Section \ref{sec:layer1}). Previously, marker lamps have been recommended by SAE~\cite{saeJ3134_201905Automated}. Most recently, Mercedes-Benz became the first automaker to receive permits allowing the use of turquoise-coloured lights on the outside of the vehicle to indicate autonomous driving mode~\cite{techspotMercedesBenzDebuts}. 
 
\item[] \textbf{High-priority messages--} Conspicuousness is essential for messages of higher importance. It is advisable to avoid displaying such information visually, as it may result in drivers overlooking significant changes~\cite{lee2023investigating}. This is because drivers frequently switch focus between the road, displays, and NDRTs~\cite{kunze2019automation}. High-priority messages might be presented through multimodal interfaces to capture the user's attention as uncertainties escalate, e.g., when a takeover becomes more probable~\cite{naujoks2019towards,zhang2023keeping}. As shown in Section~\ref{sec:multion}, Audible verbal messages accompanied by visual cues have been proven to be more effective than relying solely on visual signals when communicating takeover requests~\cite{colley2021investigating}. However, they may not be sufficient in situations where the driver is listening to music, conversing with passengers, or dealing with loud noises~\cite{miksik2020building}. As discussed in Section~\ref{sec:multion}, auditory and vibrotactile alerts could be utilised to adapt to the urgency of the message. Regardless of the modality, the warning messages should direct the user's attention towards the source of danger and persuade them to take action promptly. In regard to external stakeholders, high-priority messages involve situations ranging from intent communication to warning signals~\cite{verma2019pedestrians,faas2021pedestrian,li2018cross}. Such explanations should take into account the diverse road user requirements, such as those with impairments, language barriers, or who may be distracted. 

\end{description}

\subsubsection{Comprehensibility}
Comprehensibility is the capacity of an explanatory agent to represent its knowledge in a human-understandable way~\cite{hoffman2018metrics}. For the message to be comprehensible, it must be easily understood with minimal cognitive effort. This property highly depends on the audience, and the context since comprehensibility is a subjective concept. We have gathered four factors that support comprehensibility:

\begin{description}

\item[] \textbf{Clarity--} refers to the degree to which the resulting explanation is explicit. This property is particularly relevant for high-priority messages. The interface should adopt commonly used icons and filter out irrelevant traffic features to mitigate the risk of ambiguity~\cite{liu2023literature}. Additionally, explanations should refrain from utilizing highly abstracted information for situation awareness~\cite{colley2021investigating}. As highlighted in Section~\ref{sec:layer2}, AV needs to communicate its intentions explicitly to external stakeholders, leaving no room for interpretation. Moreover, it is crucial to avoid confusing or misleading those who are not the intended recipients. For instance, a passenger does not need to understand the information that is intended for the driver. Similarly, if the message targets pedestrians, it should not perplex other drivers who may happen to see it. 
 
\item[] \textbf{Selection and refinement--} pertain to the capacity of an explanatory agent to focus solely on the critical causes that are sufficient to explain the situation. Typically, humans do not expect an explanation to contain the complete list of causes of a decision but rather seek an explanation that conveys the most critical information supporting the decision~\cite{miller2019explanation}. When presenting driving-related information, displays should offer minimal features necessary to justify the situation~\cite{naujoks2019towards}. As pointed out in Section~\ref{sec:forw}, external communication should consider non-visual, more inclusive eHMI concepts for the safety of road users who do not have access to visuals~\cite{zhanguzhinova2023communication,haimerl2022evaluation}. 
 
\item[] \textbf{Informativeness--} relates to the ability of an explanatory agent to provide relevant information to the user without causing cognitive workload~\cite{zhang2023tactical}. As per the discussion in Section \ref{sec:multion}, when there is an unusual amount of information to convey, multimodality could be a solution to not overwhelming users by loading one sensory modality.

\item[] \textbf{Simplicity/Parsimony--} refers to the complexity of the resulting explanation. A parsimonious explanation is a simple explanation. The optimal degree of parsimony may vary depending on the user. For example, novice drivers may benefit from more detailed semantic information, while experienced or elderly drivers may find a simpler message more helpful~\cite{liu2023literature}. 

\end{description}

\subsubsection{Interactivity}

Despite its significance, bidirectional interactivity remains a relatively less explored feature within the realm of explanation communication, as revealed in Table \ref{tab:communication}. Bidirectional interactivity involves gaining insight into the target audience, including their goals, needs, and preferences~\cite{chen2018situation}. By knowing the users, the interactive experience can be tailored to suit their context, behaviour, and personality. We have identified three elements to consider for an interactive explanatory agent: engagement, empathy, and anthropomorphism.

\begin{description}

\item[] \textbf{Engaging--} Designing for engagement means making it responsive and adaptive to the users' inputs, actions, and contexts. This involves an explanation method to reason about prior interactions to interpret and respond to users' follow-up questions. This is important for creating a rapport with the users, which is also beneficial for the initial trust-building process. Then, the users should be able to control the level of engagement they prefer~\cite{colley2022effects}. Regarding other road users, communication should provide a form of acknowledgement, showing the vehicle is aware of them, as described in Section~\ref{sec:forw}. This could include a mechanism for bidirectional communication by responding to gestures~\cite{colley2021investigating}. In a V2X context, messages could be sent to HMIs of other agents to enable interactive communication.

\item[] \textbf{Empathic--} Empathic design involves designing for emotion, making explanations expressive and personality-driven. This involves using action, emotion, and gesture recognition and utilising physiological inputs (e.g., thermal, olfactory, gustatory, cerebral, or cardiac signals) to synthesise data, adapt suitable modalities, and refine interactions~\cite{jansen2022design}. The explanatory agent should identify additional relevant user characteristics (e.g., age, gender, cultural background) to address empathic interface requirements. This could also be beneficial to the conceptual trust-building process (See Section~\ref{sec:conceptual}).  

\item[] \textbf{Anthropomorphic--} By leveraging anthropomorphic features such as attractiveness, personalisation, ethnicity, and facial similarity, explanatory agents can create more engaging and user-friendly experiences. Implementing elements such as voice, tone, language, humour, and emotion recognition can contribute to a human-like relatable experience~\cite{gyevnar2022cars,ruijten2018enhancing,zang2022effects}. According to the analysis in Section~\ref{sec:forw}, anthropomorphic signals such as facial expressions or pictographs could be adapted to communicate the vehicle's intention for external stakeholders~\cite{verma2019pedestrians,colley2023scalability}.

\end{description}

Implementing an explainable AV that can adjust conspicuousness based on message priority requires thorough consideration during the design and deployment phases. Moreover, the emphasis on comprehensibility and interactivity adds further complexity to the implementation process. Therefore, there are several important risks and implications associated with these tasks that should be kept in mind, as follows:

\begin{description}

\item[] \textbf{User adaptation and familiarity--} Adjusting the communication channel's conspicuousness may require users to adapt to new modes of information processing and response~\cite{bengler2020hmi}. Ensuring that users are adequately trained and familiar with the AV's communication interfaces and protocols is crucial to facilitate quick and appropriate responses to messages of varying priorities~\cite{du2019look}.
 
\item[] \textbf{Sensory overwhelm and alert fatigue--} Using multimodal communication in AVs can lead to sensory overwhelm and alert fatigue~\cite{martinez2022exploring}. Overexposure to alerts may result in habituation, causing desensitization to urgent alerts and delayed reactions in critical takeover scenarios~\cite{fu2020too}. Drawing from Section~\ref{sec:pro-non}, continuous exposure to messages may overwhelm the user or even foster overreliance on the notifications, leading to complacency and reduced engagement with the driving environment. Striking a balance between the urgency of the message, sensory tolerance, and the user's cognitive load is crucial to prevent undue stress without compromising the user's ability to maintain situational awareness.

\item[] \textbf{Ambiguity in explanations--} Explanations must align with the AV's driving behaviour. Inconsistencies between the provided explanations and the AV's actions can result in confusion---relates to both internal and external communication. This may lead users to draw wrong conclusions or take incorrect actions. To avoid confusion and misunderstanding of the system's status, various visual feedback strategies and symbols have been proposed~\cite{dey2020taming,naujoks2019towards}. Nonetheless, further research is required to explore the use of multimodal communication, especially in high-pressure or noisy driving conditions, to determine if it contributes to ambiguity.
 
\item[] \textbf{User diversity--} As discussed in Section~\ref{sec:conceptual}, there are differences in how people of different ages, genders, and cultures perceive and trust AV technology~\cite{zoellick2019amused,qu2019development,edelmann2021cross,liu2019effect}. Further research requires consideration of these factors, along with digital literacy, cognitive abilities, etc., to ensure the information is easily understandable and identifiable by the intended stakeholders---to whom it was addressed. 
 
\item[] \textbf{Technological reliability and redundancy--} Depending on the communication channel's reliability, there may be a risk of technical failures or malfunctions, leading to a lack of timely and effective message delivery~\cite{hollander2019overtrust,faas2021pedestrian}. Implementing robust redundancy mechanisms and regular system checks are essential to minimize the risk of communication failures and ensure that critical messages reach the user promptly and reliably. 
 
\item[] \textbf{Empathic and anthropomorphic design challenges--} Incorporating empathic design elements that rely on accurate emotion and gesture recognition presents technical challenges, emphasizing the need for reliable systems to prevent misinterpretation of user emotions~\cite{lorente2021explaining,li2021spontaneous}. Additionally, implementing anthropomorphic features to improve the user experience can raise ethical and psychological concerns related to humanizing the AV system. This necessitates a balance between creating relatable experiences and avoiding potential user discomfort or confusion~\cite{waytz2014mind}.

\end{description}

The literature reviewed in Section~\ref{sec:info} and \ref{sec:com} shows that most user-centric approaches have actively engaged with human-subject studies to understand potential users' expectations and concerns regarding the explanation generation process~\cite{omeiza2023effects,chen2023adding,zhang2023keeping,colley2023scalability,avetisyan2023investigating}. Moreover, engaging with community representatives and advocacy groups can help understand local communities' specific needs and preferences. Their guidance can inform the implementation of inclusive design practices and accessible communication modalities for a broad spectrum of users. Collaborating with multidisciplinary experts also plays a pivotal role in optimizing the AV interface design. Human factors experts, cognitive psychologists, UX designers, and HCI specialists can contribute to developing effective communication interfaces in AVs~\cite{ebel2023forces,schneider2021explain,dey2020taming}. 

Other significant bodies to engage with include driver training and transportation safety organizations. They can facilitate the development of effective training programs and guidelines for users to familiarize themselves with the AV's communication interfaces and response protocols. Consultation with transportation safety regulators and authorities is necessary to ensure that the communication features align with existing safety standards and guidelines. Other regulatory authorities and standards organizations also play a significant role in ensuring compliance with industry regulations and safety standards for AV communication systems.

\subsection{Continuous learning and improvement}

The review provided in Section~\ref{sec:task}-\ref{sec:com} lacked discussions on how existing explanation concepts address the ongoing evolution of AV systems alongside the changing user needs. A continuous learning mechanism is necessary to ensure the system's explanation generation improves over time. User feedback integration, self-adaptive strategies, and incremental updates are some essential aspects to take into account for continuous learning and improvement.

\begin{description}
    
\item[] \textbf{Feedback integration--} The continuous learning process involves integrating user feedback into the learning process. By understanding user responses, concerns, and suggestions, the system can adjust its explanation strategies to meet users' evolving needs and preferences and enhance the system's overall performance~\cite{graefe2022human,barker2023feedbacklog}. Recent advancements in integrating large language models (LLMs) into autonomous vehicles set a paradigm shift, demonstrating the potential for continuous learning and personalized engagement~\cite{cui2024survey}. Thereby, users can enjoy a more seamless and intuitive interaction with AVs.

\item[] \textbf{Self-adaptive strategies--} As discussed in Section~\ref{sec:conceptual}, sociocultural context conditions human cognition and perception, and thus, the explanation needs of users may differ across cultures~\cite{kopecka2020explainable}. As AVs traverse cultural boundaries, they may need to adjust their behaviour to align with the norms and expectations of each culture encountered~\cite{pillai2017virtual,messner2022improving}. With continuous learning, the system would adapt strategies in response to changing environments by monitoring shifts in user behaviour and updating knowledge with fresh data points in real-time.

\item[] \textbf{Incremental updates and upgrades--} Continuous learning also involves the integration of incremental updates to the system's architecture, algorithms, explanation techniques, and data processing~\cite{lesort2020continual}. By doing so, the system remains up to date with the latest advancements in the field and can leverage state-of-the-art techniques to improve its performance and capabilities. However, these enhancements may not be available immediately, as manufacturers typically release updates and upgrades at set intervals, ranging from several months to a year.   

\end{description}

Integrating continuous learning mechanisms in an explainable AV system has several benefits but also poses implications related to data processing and model performance. Collecting and analysing user feedback data may raise concerns about privacy and security. Ensuring robust data encryption, anonymisation techniques, and secure data storage protocols are essential to protect sensitive user information from unauthorised access and misuse~\cite{sun2021survey}. Related to this, regular auditing of the learning models and data sources is important to identify and mitigate any algorithmic bias resulting from continuous learning~\cite{bae2022discovering,lim2019algorithmic,khan2023autonomous}. On another note, continuous learning may impact trust if users perceive the system's decisions as unpredictable or inconsistent. Providing clear and transparent explanations for the system's continuous learning processes and decision-making criteria can help users build a better mental model of the system~\cite{du2018voice,garcia2018explainable}. Concerning the model performance, continuous learning may introduce system instability, leading to unexpected behaviours or unintended consequences that could compromise user safety and trust. Implementing rigorous testing and validation procedures, along with effective fail-safe mechanisms, is crucial to ensure the stability and reliability of the AV system throughout the continuous learning process~\cite{koopman2018practical}. Finally, it is worth noting that engaging with relevant stakeholders, including data privacy and security experts, regulatory authorities, and industry specialists, is important to mitigate the potential risks and implications.
 
\section{Conclusions} 
\label{sec:conc}

In this study, we conducted a literature review to assess the state-of-the-art research in explainable autonomous driving. The study addressed several research questions related to the explanatory task, information, and means of communication. Our analysis revealed that several factors, including stakeholders, driving operations, and level of autonomy, influence the explanatory task. We discussed situations requiring explanations by contrasting proactive and reactive explanations based on their criticality—either situations that pose an immediate danger or harmless but questionable driving behaviours. Regarding explanatory information, we identified three layers of transparency: goal, reasoning, and projection, elucidating them further based on the level of autonomy and stakeholder needs. Finally, we evaluated how the explanatory information is conveyed to internal and external stakeholders. We highlighted three critical design considerations: conspicuousness, comprehensibility, and interactivity. While current approaches facilitate one-way interaction, our analysis underscores the need for further research into bidirectional interactivity, allowing users to actively engage with the system and provide feedback. Considering the lack of a systematic design approach, we suggested a roadmap to address all key design elements. By considering responsible research and innovation principles, our roadmap emphasises the importance of understanding diverse requirements for explanations. It pinpoints critical design elements related to explanatory tasks, information, and communication, as identified in the review: understanding the audience, generating timely explanations and communicating human-friendly explanations. Notably, our roadmap integrates continuous learning aspects, a dimension often overlooked in existing literature. The roadmap concerns both the internal and the external stakeholders; therefore, the suggestions can be adapted and applied to the research relating to both areas.

\section*{Acknowledgments}
We would like to thank Daniel Omeiza for his valuable feedback on the manuscript. This work was supported by the Wallenberg AI, Autonomous Systems and Software Program (WASP), funded by the Knut and Alice Wallenberg Foundation.

\bibliographystyle{named}
\bibliography{ijcai19}

\end{document}